\documentclass[journal,twoside,web]{ieeecolor}
\usepackage{generic}
\usepackage{cite}
\usepackage{amsmath,amssymb,amsfonts}
\usepackage{algorithmic}
\usepackage{graphicx}
\usepackage{algorithm,algorithmic}
\usepackage{hyperref}
\hypersetup{hidelinks=true}
\usepackage{textcomp}

\def\liminf{\lim_{t \to \infty}}

\def\L2{{\cal L}_2}
\def\L2e{{\cal L}_{2e}}

\def\rea{\mathbb{R}}

\def\diag{\mbox{diag}}

\def\qed{\hfill$\Box \Box \Box$}
\def\calk{{\cal K}}
\def\caly{{\cal Y}}

\def\calc{{\cal C}}

\def\cala{{\cal A}}
\def\calf{{\cal F}}

\def\liminf{\lim_{t \to \infty}}

\def\L2e{{\cal L}_{2e}}

\def\rea{\mathbb{R}}
\def\intnum{\mathbb{Z}}

\def\diag{\mbox{diag}}

\def\col{\mbox{col}}

\def\diag{\mbox{diag}}
\def\rank{\mbox{rank}\;}

\def\begsubequ{\begin{subequations}}
	\def\endsubequ{\end{subequations}}
\def\begpro{\begin{proposition}}
	\def\endpro{\end{proposition}}
\def\beglem{\begin{lemma}}
	\def\endlem{\end{lemma}}
\def\begass{\begin{assumption}}
	\def\endass{\end{assumption}}
\def\begcor{\begin{corollary}}
	\def\endcor{\end{corollary}}
\def\begproo{\begin{proof}}
	\def\endproo{\end{proof}}

\def\begmat#1{\begin{bmatrix}#1\end{bmatrix}}
\def\begali#1{\begin{align}{#1}\end{align}}
\def\begalis#1{\begin{align*}{#1}\end{align*}}

\newtheorem{lemma}{Lemma}
\newtheorem{proposition}{Proposition}
\newtheorem{corollary}{Corollary}
\newtheorem{fact}{Fact}

\newtheorem{remark}{Remark}

\newtheorem{assumption}{Assumption}

\def\lab{\label}
\def\begequ{\begin{equation}}
	\def\endequ{\end{equation}}
\def\begrem{\begin{remark}\rm}
	\def\endrem{\end{remark}}

\def\BibTeX{{\rm B\kern-.05em{\sc i\kern-.025em b}\kern-.08em
    T\kern-.1667em\lower.7ex\hbox{E}\kern-.125emX}}
\begin{document}
\title{Adaptive State Observers of Linear Time-varying Descriptor Systems:\\ A Parameter Estimation-Based Approach}
\author{Romeo Ortega, \IEEEmembership{Life Fellow, IEEE}, Alexey Bobtsov, Fernando Castaños, \IEEEmembership{Member, IEEE} and Nikolay Nikolaev
	\thanks{Manuscript received XXX; accepted XXX.
		Date of publication XXX; date of current version XXX. The work was written with the financial support of the Russian Federation (project no. FSER-2025-0002).  This work was also supported by 111 project, No. D17019. (Corresponding author:
		Alexey Bobtsov.)}
	\thanks{	Romeo Ortega is with Department of Electrical and Electronic Engineering, ITAM, Ciudad de M\'exico, M\'{e}xico. (e-mail: romeo.ortega@itam.mx)}
	\thanks{	Alexey Bobtsov is with the Hangzhou Dianzi University ITMO Joint Institute, Xiasha Higher Education Zone, Hangzhou, Zhejiang, People’s Republic of China and Faculty of Control Systems and Robotics, ITMO University, Kronverkskiy av. 49, St. Petersburg, 197101, Russia (e-mail: bobtsov@mail.ru)}
	\thanks{	Fernando Castaños is with the Departamento de Control Autom\'atico, Cinvestav, IPN, M\'exico. 	(e-mail: castanos@ieee.org)}
	\thanks{	Nikolay Nikolaev is with the Faculty of Control
		Systems and Robotics, ITMO University, Kronverkskiy av. 49, St. Petersburg, 197101, Russia (nikona@yandex.ru)}
}

\maketitle

\begin{abstract}
In this paper, we apply the recently developed generalized parameter estimation-based observer design technique for state-affine systems to the practically important case of linear time-varying descriptor systems with uncertain parameters.  We give simulation results of benchmark examples that illustrate the performance of the proposed adaptive observer.
\end{abstract}


\section{Introduction}
\lab{sec1}
%
One of the most elegant and simple ways to model a physical system is the creation of separate models for standardized subcomponents that can then be pasted together via a network---this procedure leads naturally to  models consisting of differential and  algebraic equations. These systems are called \emph{descriptor systems}, singular systems, or differential-algebraic (DAE) systems. The modeling concept mentioned above is used in many modern CAD/modeling systems like SIMULINK, Scicos, MODELICA, and DYMOLA which are applicable to multi-physics problems from different physical domains including mechanical, mechatronic, fluidic, thermic, hydraulic, pneumatic, elastic, plastic or electric components, see \cite{ILCREIbook, MEHUNG} for further details. 
In most modeling systems reduction to an explicit model, if at all possible, is performed  with the accompanying loss of accuracy and sparsity---therefore, it is convenient to preserve the DAE structure. Designing control laws for these systems typically requires the knowledge of their state variables. This information is also required in other applications like fault detection and isolation \cite{PATetalbook}. A lot of research has been reported in the last few years on the development of {\em state observers} for this kind of systems, see {\em {\em e.g.}}, \cite{BOBCAM, DAIbook, DARscl12, HOUMUL, MULHOU}. In this paper we are particularly interested in the case of {\em linear time-varying} (LTV) descriptor systems---see \cite{CAMDELNIK} for an early reference on this topic and \cite{BOBCAM,ZETetal} for more recent references.    

Although there has been a considerable amount of research done on the design of observers for linear time-invariant (LTI) descriptor systems, there has been very little work on the design of observers for LTV ones. This, in spite of the fact that LTV descriptor systems naturally arise from the linearization along a trajectory of nonlinear systems---with the additional difficulty that the linear approximation contains uncertain terms due to the fact that the chosen trajectory is not necessarily a solution of the original system, leading to the presence of {\em uncertain parameters} in the DAE model of the system, see \cite{CAM} for details. In view of the latter reason we are interested here in the design of {\em adaptive} observers. 

Much of the existing literature on this topic makes extensive use of nonlinear coordinate changes and differentiations of computed quantities, in particular the input signal---a process called {\em completion} \cite{BOBCAM}. While this is theoretically attractive, differentiating signals is not practical due to the presence of noise. It is sometimes argued, that there are some cases, for instance in index three DAE from constrained mechanics, that the derivative of the input signal does not appear in the observer  \cite{BOB}. But it is, of course, of interest to develop a procedure to design observers that does not rely on these non-robust operations. Similarly, we are interested here in the design of (what is called) {\em normal} observers \cite[Chapter 4]{DAIbook}, which are described by ordinary differential equations. This, in contrast with {\em singular} observers, that are themselves described by DAE, that may exhibit undesirable impulsive behavior \cite{DAIbook,KUNMEHbook}. 

The approach adopted in the paper for the design of our adaptive observers relies on the use of {\em generalized parameter estimation-based observers} (GPEBO) recently reported in \cite{ORTetalaut}. The main feature of GPEBO is that the problem of state observation is recasted as a problem of {\em parameter} estimation, namely of the systems initial condition. This approach has been proven very successful for the state observation of state-affine systems, which are recast as LTV systems, and many extensions and practical applications to the method have been reported \cite{BEZetal,BOBetalijc,BOBetalaut,PYRetal,ROMORT}. In a recent paper \cite{WANORTBOB} it was shown that state observation of LTV systems is possible imposing only the (necessary) assumption of observability---a result that should be contrasted with the usual (far stronger) uniform complete observability assumption required by the standard Kalman-Bucy filter solution \cite{BERbook,RUGbook}.\footnote{A related property of interest for LTV systems is {\em reconstructibility}, which is equivalent to complete controllability of the dual system \cite[Proposition 3.5]{ILC}.}   Our main objective in this paper is to prove that adopting the GPEBO approach it is possible to generate new, stronger, solutions to the problem of design of adaptive observers for LTV descriptor systems. 

Following \cite[Section 3]{KUNMEHbook}, the differences between the LTI and the LTV cases will be underscored in the paper. In particular, we show  that assumptions that are always invoked in the LTI case do no play any role in the LTV case. For instance, we show that {\em regularity}---a key assumption for LTI systems implied by Conditions A1-A3 of \cite[Subsection 2.2.2]{BOBCAM}---is irrelevant in the LTV case  since, as shown in  \cite[Subsection 3.1]{KUNMEHbook}, for LTV implicit systems (pointwise) regularity and solvability of the equations are {\em unrelated} properties. Another important property of LTV systems is the so-called {\em strangeness index} \cite[Definition 3.15]{KUNMEHbook}, which in LTI systems is called the nilpotency index, that plays a central role in the observer design. Indeed, we show that if this index is one then designing a GPEBO is a trivial task. We prove that developing a general theory for observer design in the case of LTV descriptor systems with index larger than one is far from trivial, and some particular solutions are given in that case.

The rest of the paper has the following structure. As an introduction to the general GPEBO theory, in Section \ref{sec2} we design an adaptive observer for a benchmark example applying GPEBO. In Section \ref{sec3} we invoke the Standard Canonical Form \cite{CAMPET} to develop a general framework for observer design, and solve some specific cases. Section \ref{sec4} is devoted to the solution using GPEBO of a benchmark  circuit example reported in \cite{BOB,BOBCAM}, which is described by a descriptor, LTV system. Finally, Section \ref{sec5} presents some concluding remarks. A preliminary basic lemma used in the GPEBO design is given in Appendix \ref{appa}.\\

\noindent {\bf Notation.} $I_n$ is the $n \times n$ identity matrix and  ${0}_{s \times r}$ is an $s \times r$ matrix of zeros. $\rea_+$ and $\intnum_+$ denote the positive real and integer numbers, respectively. For $q \in \intnum_+$ we define the set $\bar q:=\{1,2,\dots,q\}$. For a vector $a \in \rea^n$, we denote its Euclidean norm by $|a|$ and for a matrix $A \in \rea^{n \times m}$ its Euclidean matrix norm is denoted $\|A\|$. The action of {a linear time-invariant (LTI)} filter $\calf(p) \in \rea(p)$ on a signal $w(t)$ is denoted as $\calf(p)[w]$, where $p^n[w]:={d^n w(t)\over dt^n}$. For all time-varying matrices $M:\rea_+ \to \rea^{m \times n}$ all properties ({\em e.g.}, rank) are defined pointwise in time. {To avoid dealing with delicate theoretical technical issues related to signals smoothness, we assume that all functions are sufficiently smooth (or analytic) and all differential equations {\em analytically solvable}---that is, solutions exist and they are determined by the state initial conditions.} 
%

\section{An Introduction to GPEBO}
\lab{sec2}
In this section we illustrate the GPEBO design with a classical example of an LTV descriptor system with uncertain parameters.
\subsection{A motivating example}
\lab{subsec21}
%
In this section we consider the following LTV descriptor system\footnote{In the LTI case, this is called {\em second equivalent form} in \cite[Subsection 1-3.2]{DAIbook}.} with constant uncertain parameters
\begsubequ
\lab{oldsys}
\begali{
	\lab{oldsysa}
	E\dot x&=A(t)x+B(t) u+F(t)\theta \\
	\lab{oldsysb}
	y& =C(t) x,
} 
\endsubequ
where we  defined the matrices
\begali{
	\nonumber
	E&:=\begmat{I_{n_a} & 0_{n_a \times n_b} \\ 0_{n_b\times n_a} & 0_{n_b \times n_b}}, A(t):=\begmat{A_{11}(t) & A_{12} (t)\\ A_{21}(t) & A_{22}(t)},\\
	\lab{oldsysc}
	B(t)&:=\begmat{B_1(t) \\ B_2(t)},
	F(t):=\begmat{F_1(t) \\ F_2(t)}, C(t) :=\begmat{C_a(t) \\ C_b(t)}^\top,
} 
and partition the state as $x:=\begmat{x_a \\ x_b}$, with $x_a(t)\in \rea^{n_a}$, $x_b(t)\in \rea^{n_b}$, $n:=n_a+n_b$, $u(t) \in \rea^m$, $y(t) \in \rea^r$, $r < n$, $A(t)\in \rea^{n \times n},B(t)\in \rea^{n \times m},C(t)\in \rea^{n \times r}$ and $F(t)\in \rea^{n \times q}$ are  {\em known} functions of time, and $\theta \in \rea^q$ is an {\em unknown} constant vector.  Without loss of generality, throughout the paper we assume that $C(t)$ is {\em full rank}. 

For the design of the adaptive GPEBO we impose the following.

\begin{assumption} 
	\lab{ass1}\em
	The matrix 
	
	$\cala_b(t):=\begmat{ A_{22}(t) \\ C_b(t)} \in \rea^{ (n_b+r)\times n_b }$ 
	
	is {\em full rank and injective}.\footnote{This assumption is called  {\em impulse observability} (or observability at $\infty$) in the LTI case, see {\em e.g.} \cite{BunBMN99,DAIbook}.}
\end{assumption}

Equipped with this assumption we can state the following. 

\begin{proposition} \em 
	\lab{pro1}
	Consider the descriptor LTV system \eqref{oldsys}, \eqref{oldsysc} verifying { Assumption \ref{ass1}}. There exists an adaptive GPEBO  of the form 
	\begali{
		\nonumber
		\dot{\chi} & =  F( \chi,y, u)\\
		\nonumber 
		\hat x &  =  H(\chi,y,u)\\
		\lab{obssys}
		\hat \theta &  =  M(\chi,y,u)
	}
	with $\chi(t) \in \rea^{n_\chi}$ and the mappings
	$$
	F:\rea^{n_\chi \times r \times m} \to \rea^{n_\chi},\;H:\rea^{n_\chi \times r \times m} \to \rea^{n},\;M:\rea^{n_\chi \times r \times m} \to \rea^{q},
	$$
	such that, for all initial conditions, we have that 
	\begequ
	\lab{concon}
	\liminf \Big|\begmat{\hat x(t)-x(t) \\ \hat \theta(t)-\theta} \Big|=0,
	\endequ
	{\em exponentially fast}, provided some suitable {\em excitation} conditions---stated in { Assumption \ref{ass2}} below---are satisfied.
\end{proposition}

\begproo
	First, notice that from the definition of $y$ and $C(t)$ in \eqref{oldsysb} and \eqref{oldsysc}, respectively, and the last $n_b$ rows of \eqref{oldsysc} we have the following relations
\begin{align}
	\nonumber
\cala_b(t) x_b&=\begmat{ A_{22}(t) \\ C_b(t)}x_b=-\begmat{A_{21}(t) \\ C_a(t)}x_a-\begmat{B_2(t) \\ 0_{n_b \times m}}u \\
\nonumber
	&-\begmat{F_2(t) \\ 0_{n_b \times q}}\theta+\begmat{0_{n_a \times 1} \\ y}.
\end{align}

	On the other hand, {Assumption 1} ensures the existence of the Moore-Penrose pseudoinverse of $\cala_b(t)$ defined as
	$$
	\cala^\dagger_b(t):=[\cala_b^\top(t) \cala_b(t)]^{-1} \cala^\top_b(t),
	$$ 
	see {\em e.g.}  \cite{CamM09}, from which we get
	\begin{align}
	\lab{xb}
	\nonumber
	x_b=&-\cala_{b}^\dagger(t)\Bigg[\begmat{A_{21}(t) \\ C_a(t)}x_a  +\begmat{B_2(t) \\ 0_{n_b \times m}}u + \begmat{F_2(t) \\ 0_{n_b \times q}}\theta \Bigg] \\
	&+\cala_{b}^\dagger(t)\Bigg[\begmat{0_{n_a \times 1} \\ y}\Bigg].
	\end{align}
	Hence, the  first $n_a$ rows of \eqref{oldsysc} take the form
	\begalis{
		&\dot x_a =A_{11}(t) x_a-A_{12}(t)\cala_{b}^\dagger(t)\Big[\begmat{A_{21}(t) \\ C_a(t)}x_a+\begmat{B_2(t) \\ 0_{n_b \times m}}u \Big]\\
		&-A_{12}(t)\cala_{b}^\dagger(t)\Big[\begmat{F_2(t) \\ 0_{n_b \times q}}\theta	- \begmat{0_{n_a \times 1} \\ y}\Big] +B_1(t) u+F_1(t) \theta\\
		&=:A_0(t) x_a+B_0(t) u+D_0(t) \theta +A_{12}(t)\cala_{b}^\dagger(t) \begmat{0_{n_a \times 1} \\ y},
	}
	where we defined the matrices
	\begalis{
		A_0(t) &:=A_{11}(t)-A_{12}(t)\cala_{b}^\dagger(t)\begmat{A_{21}(t) \\ C_a(t)},\\
		B_0(t)&:=B_1(t)-A_{12}(t)\cala_{b}^\dagger(t)\begmat{B_2(t) \\ 0_{n_b \times m}},\\
		D_0(t) &:= F_1(t)-A_{12}\cala_{b}^\dagger\begmat{F_2(t) \\ 0_{n_b \times q}}.
	}
	Let us introduce the {\em dynamic extension}
	\begalis{
		\dot \xi_a&= A_0 (t)\xi_a+ B_0(t) u +A_{12}(t)\cala_{b}^\dagger(t) \begmat{0_{n_a \times 1} \\ y}\\
		\dot \xi_b &= A_0(t) \xi_b+ D_0(t)\\
		\dot \Phi&=A_0(t) \Phi,\;\Phi(0)=I_{n_a}.
	}
	Define the error signal
	\begin{equation} \label{error_signal}
		e:=\xi_a+\xi_b \theta-x_a,
	\end{equation}
	which satisfies the homogeneous differential equation
	$$
	\dot e = A_0(t) e.
	$$
	Using the property of its fundamental matrix $\Phi$ given in Appendix \ref{appa} we can write 
	$$
	e(t)=\Phi(t)e_0
	$$
	where $e_0=:e(0)\in \rea^{n_a}$ is an {\em unknown} parameter vector. Replacing this into \eqref{error_signal}  we get 
	\begequ
	\lab{xa}
	x_a=\xi_a+\xi_b \theta-\Phi e_0.
	\endequ
	From \eqref{xa} it is clear that if the parameters $\theta$ and $e_0$ are known then we can compute the state $x$. This motivates us to define an {\em estimate} of the state $x_a$ as
	$$
	\hat x_a=\xi_a+\xi_b \hat \theta-\Phi \hat e_0,
	$$
	where $\hat \theta,\hat e_0$ are estimates of the unknown parameters $\theta$ and $e_0$, respectively. It is clear that if $\hat \theta(t) \to \theta$ and $\hat e_0(t) \to e_0$ then $|\hat x(t)-x(t)| \to 0$. 
	
	Our remaining task is then to propose a parameter estimator for  $\theta$ and $e_0$. Towards this end, it is necessary to defined a function where the parameters are related with {\em measurable} signals, such a relation is called a {\em regression equation} \cite{SASBODbook,TAObook} and is generated as follows. From  the definition of $y$ \eqref{oldsysb}, \eqref{oldsysc} we get
	\begalis{
		&y=C_a(t)x_a+C_b(t) x_b\\
		&=C_a(t)(\xi_a+\xi_b \theta-\Phi e_0)+C_b(t) x_b\\
		&=C_a(t)(\xi_a+\xi_b \theta-\Phi e_0)-C_b(t)\cala_{b}^\dagger\Big[-\begmat{0_{n_a \times 1} \\ y}\Big]\\
		&-C_b(t)\cala_{b}^\dagger\Big[\begmat{\begmat{A_{21} (t)\\ C_a(t)}x_a + B_2(t) \\ 0_{n_b \times m}}u+\begmat{F_2(t) \\ 0_{n_b \times q}}\theta\Big]\\
		&=C_a(t)(\xi_a+\xi_b \theta-\Phi e_0)-C_b(t)\cala_{b}^\dagger(t)\Big[{-\begmat{0_{n_a \times 1} \\ y}} \Big]\\
		&-C_b(t)\cala_{b}^\dagger(t)\Big[ \begmat{A_{21}(t) \\ C_a(t)}\Big(\xi_a+\xi_b \theta-\Phi e_0 \Big)\Big]\\
		&-C_b(t)\cala_{b}^\dagger(t)\Big[\begmat{B_2(t) \\ 0_{n_b \times m}}u+\begmat{F_2(t) \\ 0_{n_b \times q}}\theta\Big].
	}
	where we used \eqref{xa} to get the second identity, \eqref{xb} in the third one, and  \eqref{xa} again in the last one. Grouping on the left hand side {\em measurable} terms and on the right hand side terms depending on the unknown parameters $\eta:=\begmat{ \theta \\  e_0} \in \rea^{q+n_a}$ we get a  {\em linear regression equation} (LRE) of the form 
	\begequ
	\lab{lre}
	Y=\psi \eta,
	\endequ
	where we have defined the measurable signals
	\begali{
		\nonumber
		Y &:=y-C_a(t)\xi_a -C_b(t)\cala_{b}^\dagger (t)\begmat{0_{n_a \times n_p} \\ y}\\
		\nonumber
		& +C_b(t)\cala_{b}^\dagger(t)\begmat{A_{21}(t) \\ C_a(t)}\xi_a+C_b(t) \cala_{b}^\dagger(t)\begmat{B_2(t) \\ 0_{n_b \times m}}u \\
		\lab{regpsi}
		\psi &:=\begmat{\psi_1 & \vdots & \psi_2},
	}
	with
	\begin{footnotesize}
	\begin{align*}
		\psi_1&= \Big(C_a(t) -C_b(t)\cala_{b}^\dagger(t) \begmat{A_{21}(t) \\ C_a(t)}\Big)\xi_b -C_b(t)\cala_{b}^\dagger(t)\begmat{F_2(t) \\ 0_{n_b \times q}},\\
		\psi_2&= \Big[C_b(t)\cala_{b}^\dagger(t)\begmat{A_{21}(t) \\ C_a(t)}-C_a(t)\Big]\Phi,
	\end{align*}
\end{footnotesize}
	with $Y(t) \in \rea^{r}$ and $\psi(t)\in \rea^{r \times (q+n_a)}$.
	
	To estimate the parameters from the LRE \eqref{lre} we propose to use an advanced estimator, see discussion point {\bf D4} below for further detail on this issue. For instance, we can  use the least-squares plus dynamic regressor extension and mixing parameter  estimator proposed in \cite{ORTROMARA}, which ensures \eqref{concon} provided that the regressor $\psi$ is {\em interval exciting} (IE) \cite{KRERIE,TAObook}. 
	
	Thus, to complete the proof of the proposition we make the following.
	
\begin{assumption}  
		\lab{ass2}\em
		The regressor $\psi(t)$ is IE, {\em i.e.}, there exists constants $t_c>0$ and $\rho>0$ such that
		$$
		\int_0^{t_c}\psi(s)\psi^\top(s)ds \geq \rho I_r.
		$$ 
	\end{assumption}
\endproo
\subsection{Discussion}
\lab{subsec22}
%
\noindent {\bf D1} [{\em Non adaptive observer}] In the {\em absence} of uncertain parameters, {\em i.e.}, if $F(t) \equiv 0_{n \times q} $, the observer design can be significantly simplified. Indeed,  in that case we can omit the states $\xi_b$ and define the error signal as   $e:=\xi_a-x_a$ to obtain the linear regression equation $Y=\psi_0 e_0$ with the new (reduced) regressor
\begequ
\lab{psi0}
\psi_0:= -\Big[C_a(t)-C_b(t)\cala_{b}^\dagger(t)\begmat{A_{21}t) \\ C_a(t)}\Big]\Phi.
\endequ

\noindent {\bf D2} [{\em About the assumption on $\cala_b(t)$}] In the {\em LTI case}, it is possible to show that {Assumption 1} is {\em equivalent} to the condition
$$
\rank\Bigg\{\begmat{E & A\\0_{p\times n} & C\\0_{n \times n} & E}\Bigg\}=\rank\{E\}+n,
$$ 
which is a necessary and sufficient condition for {\em impulse observability}, see \cite[Definition 2.3.3]{DAIbook} for a definition of the latter and  \cite[Theorem 2.3.4]{DAIbook} for the proof of the claim. This assumption is ubiquitous in most papers on observers for implicit LTI systems, see  \cite[Theorem 1]{MULHOU} for a relaxation of this condition involving the matrix $B$. In the LTV case, {Assumption 1} is related to the fact whether an output feedback exists that allows to make the system regular and strangeness-free---see Subsection \ref{subsec41} for a discussion on strangeness-free systems and \cite{KUNMEHbook,KunMR01} for an in-depth discussion on the topic.
\\

\noindent {\bf D3} [{\em About the IE assumption}] In the LTI case, the IE {Assumption 2} is intimately related with observability of the system. As explained in {\bf D1} above, in the {\em absence} of uncertain parameters we require IE of the simplified regressor $\psi_0$ given in \eqref{psi0}. If, moreover $C_b=0_{r \times n_b}$ then the regressor reduces to $-C_a \Phi$ and $A_0$ becomes
$$
A_0= A_{11}-A_{12}A_{22}^{-1}\begmat{A_{21} \\ C_a}.
$$ 
In this case, it can be shown \cite{WANORTBOB} that this regressor is IE {\em if and only if} the pair $(A_0,C_a)$ is {\em observable}. 

Hence, even in the LTI case, the relationship between IE and observability of the original system is far from obvious. Clearly, the addition of $C_b$ and the time-varying terms increases the possibilities to verify the IE assumption.\\

\noindent {\bf D4} [{\em About the parameter estimator}] Given the LRE \eqref{lre} it is possible to implement a classical recursive parameter estimator, {\em e.g.}, gradient or least-squares \cite{SASBODbook,TAObook}. However, it is well-known that to ensure parameter convergence, these schemes impose the highly restrictive assumption of persistent excitation. Namely, that it exists a time window $T>0$ and a constant $\rho>0$ such that
$$
\int_t^{t+T}\psi(s)\psi^\top(s)ds \geq \rho I_2,\;\forall t \geq 0.
$$
It is clear that the IE assumption is significantly weaker than persistent excitation. Actually, it has been shown in \cite{WANORTBOB} that IE is a {\em necessary and sufficient} condition for the (on- or off-line) solution of the parameter estimation problem of the LRE  $Y=\psi \eta$. See  \cite{ORTetaltac} for further discussion on this issue and \cite{ORTNIKGER} for a recent survey on new parameter estimators. \\ 

\noindent {\bf D5} [{\em Possible extensions}] The assumption on $E$ can be relaxed, we only need to be able to express $x_b$ as a function of $x_a,u$ and $\theta$, as done in  \eqref{xb}. Also, notice that the result can be easily extended to systems with {\em delays} \cite{BEZetal} and external disturbances with {\em uncertain} internal model \cite{BOBetalijc,PYRetal}.
%
\section{The Standard Canonical Form}
\lab{sec3}
%
One of the main contributions of the paper is the design of a GPEBO  for regular, LTV  descriptor systems, which are transformed to the  {\em standard canonical} form \cite{CAMPET}. In this section we present this form and introduce an alternative representation for part of the dynamics that is instrumental for the design of the GPEBO, which is carried-out in the next section. For the sake of simplicity, we consider the case of systems {\em without} uncertain parameters, in the understanding that---as explained in point {\bf D1} of the previous section---this additional feature can be easily accommodated in the observer design. 
\subsection{Classical representation of the Standard Canonical Form}
\lab{subsec31}
The lemma below gives a classical representation of the descriptor system, known as the Standard Canonical form  \cite{CAMPET}.

\begin{lemma}\em \cite[Subsection 2.6]{TRE}
	Consider the {analytically solvable} {\em LTV descriptor} system 
	\begsubequ
	\lab{sys}
	\begali{
		\lab{sysx}
		E(t)\dot x(t)&=A(t)x(t)+f(t)\\
		y &=C(t)x
	}
	\endsubequ 
	with $x(t) \in \rea^n$, $y(t) \in \rea^r$, $f(t) \in \rea^n$, $E(t)\in\rea^{n \times n}, A(t)\in\rea^{n \times n}, C(t)\in\rea^{r \times n}$. There exists full rank, analytic, square matrices $P(t)\in\rea^{n \times n}$ and $Q(t)\in\rea^{n \times n}$ such that 
	\begalis{
		Q(t)E(t)P(t)&=\diag\{I_{n_a},N(t)\}\\
		Q(t)A(t)P(t)-Q(t)E(t)\dot P(t)&=\diag\{A_a(t),I_{n_b}\}
	}
	with $n_a+n_b=n$, $A_a(t) \in \rea^{n_a\times n_a}$ and $N(t)\in \rea^{n_b\times n_b}$ is {\em strictly lower triangular}, that is,
	\begequ
	\lab{n}
	N(t)=\begmat{0 & 0 & \cdots & 0\\ n_{2,1}(t) & 0 & \cdots & 0 \\ \vdots & \vdots & \ddots & \vdots \\ n_{n_b,1}(t) & n_{n_b,2}(t) & \cdots & 0} 
	\endequ
	\qed 
	\endlem
	
	From the lemma we directly conclude that, after the change of coordinates $x=P(t)z$ and premultiplying the resulting ODE by $Q(t)$, we obtain the equivalent system representation
	\begsubequ
	\lab{weisys}
	\begali{
		\lab{weisysa}
		\dot z_a&=A_a(t) z_a+f_a(t)\\
		\lab{weisysb}
		N(t)\dot z_b&=z_b+f_b(t)\\
		\lab{weic}
		y&={\mathbb C}(t) z={C}_a(t) z_a + {C}_b(t) z_b,
	}
	\endsubequ
	where we have partitioned $z=\begmat{z_a \\ z_b}$, defined $\begmat{f_a(t) \\ f_b(t)}:=Q(t)f(t)$ and the new output matrix 
	\begequ
	\lab{matbbc}
	{\mathbb C}(t):=C(t)P(t),
	\endequ
	that we partitioned as ${\mathbb C}(t)=\begmat{C_a(t) & C_b(t)}$, with $C_a(t) \in \rea^{r\times n_a}$, $C_b(t) \in \rea^{r\times n_b}$.
	
	\begrem
	\lab{rem1}
	A similar result is available for LTI systems, where it is called {\em first equivalent form} in \cite[Subsection 1-3.1]{DAIbook} or {\em Weierstrass canonical form} \cite[Lemma 1-2.2]{DAIbook}. In this case the matrix $N$ takes the simpler form
	$$
	N=\begmat{0 & 1 & 0 & \cdots & 0 & 0\\0 & 0 & 1 & \cdots & 0 & 0\\ \vdots & \vdots & \vdots & \cdots & \vdots & \vdots \\
		0 & 0 & 0 & \cdots & 0 & 1\\0 & 0&  0 & \cdots & 0 & 0}\in \rea^{n_b \times n_b}.
	$$
	But some of the difficulties for the observer design that we describe below for the LTV case, are also present in the LTI case---hence the observer design at this level of generality---even for LTI descriptor systems---is far from obvious.
	\endrem
	\subsection{A suitable representation for the $z_b$ dynamics}
	\lab{subsec32}
	%
	It is clear from \eqref{weisysa} that the design of a GPEBO for the $z_a$ dynamics proceeds seamlessly provided a suitable assumption on the output signal \eqref{weic} is imposed. This task is carried out in the next section. On the other hand, the $z_b$ dynamics is quite complex and the design of an observer requires some non-standard developments, which are presented in the next section. As a preparation for this task it is necessary to introduce a new representation of this dynamics which is carried-out below. 
	
	Given the definition of $N(t)$ in \eqref{n}, we see that the equations for $z_b$ are of the form
	\begali{
		\nonumber
		& 0=z_{b,1}+f_{b,1}(t)\\
		\nonumber
		& N_{2,1}(t)\dot z_{b,1} =z_{b,2}+f_{b,2}(t)\\
		\nonumber 
		&\vdots \hspace{2.5cm} \vdots \\
		\nonumber
		&N_{n_b,1}(t)\dot z_{b,1}+N_{n_b,2}(t)\dot z_{b,2}+\cdots +N_{n_b,(n_b-1)}(t)\dot z_{b,(n_b-1)}\\
		&=z_{b,n_b}+f_{b,n_b}(t)
		\lab{zbsys}
	}
	Designing an observer for \eqref{zbsys} at this level of generality is a formidable task. We notice that a critical condition to solve it is that the coefficients $n_{i,j}(t)$ of the matrix $N(t)$ shouldn't be equal to zero. Indeed, let us assume that $n_{2,1}(t) \equiv 0$. Then, after the replacement of the first two equations in the third one we get
	$$
	z_{b,3}=-N_{3,1}(t)\dot f_{b,1}(t)-N_{3,2}(t)\dot f_{b,2}(t)-f_{b,3}(t),
	$$
	which cannot be computed without signal differentiation.\footnote{Here, and throughout the paper, we impose the practically reasonable condition that differentiation of signals is not admissible.} With some simple calculations, it is possible to prove that similar unsolvable scenarios appear if other coefficients of the matrix $N(t)$ are identically zero. 
	
	Therefore we make the following.
	
	\begass \em
	\lab{ass4}
	The coefficients of the matrix $N(t)$ are different from zero for all $t \geq 0$.
	\qed
	\endass
	
	Equipped with this assumption we can solve the problem of estimation of the state $z_b$. To avoid cluttering the notation, let us illustrate the procedure for the case $n_b=3$. In that case, we deal with the system
	\begali{
		\nonumber
		& 0=z_{b,1}+f_{b,1}(t)\\
		\nonumber
		& N_{2,1}(t)\dot z_{b,1} =z_{b,2}+f_{b,2}(t)\\ 
		&N_{3,1}(t)\dot z_{b,1}+N_{3,2}(t)\dot z_{b,2} =z_{b,3}+f_{b,3}(t)
		\lab{zbsys3}
	}
	
	After some simple algebraic manipulations, and invoking {Assumption \ref{ass4}}, we can write part of the system \eqref{zbsys3} in the form
	\begali{
		\lab{dotzw}
		\dot z_w &=A_w(t)z_w+d_w(t)+B_w \zeta,
	}
	where we defined
	\begin{align*}
	z_w:&=\begmat{z_{b,1} \\ z_{b,2}},\;A_w(t):=\begmat{0 & a_1(t) \\ 0 & a_2(t)},\; d_w(t):=\begmat{d_{1}(t)\\d_{2}(t)},\\
	B_w&=\begmat{0 \\ 1}.
	\end{align*}
	with the known functions
	\begalis{
		a_1(t)&:= {1 \over N_{2,1}(t)},\;d_{1}(t) := {f_{b,2}(t) \over N_{2,1}(t)}\\ 
		a_2(t)&:=-{N_{3,1}(t) \over N_{3,2}(t)},\; d_{2}(t):={f_{b,3}(t) \over N_{3,2}}-{N_{3,1}(t)\over N_{3,2}(t)}{f_{b,2}(t) \over N_{2,1}(t)}
	}
	and $\zeta(t) \in \rea$ is the {\em unknown} signal
	$$
	\zeta:={z_{b,3} \over N_{3,2}(t)}.
	$$
	
	We are, therefore, in the scenario of observation of the state of the LTV system \eqref{dotzw}, which contains an {\em unknown input} signal. Although for LTI systems there exists a large literature on the topic---see {\em e.g.} \cite{TRAetal} for a recent survey---very little literature is available for LTV systems, with existing results relying on the use of noise-sensitive sliding modes and signal differentiation \cite{DAVFRILEV}. 
	\section{Design of GPEBO for the System \eqref{weisys}}
	\lab{sec4}
	%
	In this section we proceed for the design of our GPEBO for the system \eqref{weisys}. It will be shown in Subsection \ref{subsec41} that there is a particular case where the design is trivial. In Subsection \ref{subsec42} we prove that the observation of the state $z_a$ is also straightforward via GPEBO. On the other hand, observing the state $z_b$ is very challenging and several assumptions need to be imposed. 
	\subsection{The strangeness-free case}
	\lab{subsec41}
	A case of particular interest is when $N(t)\equiv 0$, which is referred  to a {\em strangeness-free} \cite[Definition 2.8]{MEHUNG}. {See also \cite[Theorem 3.60]{KUNMEHbook} for a similar result.} Notice that in this case $z_b(t)=-f_b(t)$ and the observer design task is trivialized, as it boils down to the design of an observer for the state $z_a$ of the LTV subsystem \eqref{weisysa}. 
	
	The following result follows directly from \cite[Proposition 2]{WANORTBOB} and the derivations carried out in {Proposition} \ref{pro1} to estimate an unknown parameter vector.
	
	\begpro \em
	\label{pro2}
	Consider the LTV descriptor system \eqref{sys}. Assume the system is {\em strangeness-free} and represent the system in the Standard Canonical form \eqref{weisys} with $N(t) \equiv 0$. Assume further that
	$$
	f_a(t)=f_{0a}(t)+F_a(t) \theta,
	$$
	where $f_{0a}(t) \in \rea^{n_a}$ and $F_a(t) \in \rea^{n_a \times q}$ are {\em known} and $\theta \in \rea^{q}$ is a vector of constant {\em unknown} parameters. Assume the pair $(A_a(t),C_a(t))$ is {\em observable}. Then, there exists a GPEBO of the form
	\begali{
		\nonumber
		\dot{\chi}_a & =  F_a( \chi_a,y, f_a)\\
		\lab{hatza}
		\hat z_a &  =  H_a(\chi_a,y,f_a)
	}
	with $\chi_a(t) \in \rea^{n_{\chi_a}}$  such that, for all initial conditions, we have that 
	\begequ
	\lab{liminfa}
	\liminf |\hat z_a(t)-z_a(t)|=0,
	\endequ
	{\em exponentially fast}. 
	\endpro
	\subsection{GPEBO-based observer for $z_a$}
	\lab{subsec42}
	When the system is not strangeness-free the matrix $N(t) \neq 0$ and we cannot compute $z_b$ from the knowledge of $f_b(t)$. This, in its turn imply that we cannot generate a measurable quantity that depends linearly on $z_a$. An assumption is then required to achieve the latter objective.\footnote{To simplify the presentation we consider here the case when there are no uncertain parameters.}  
	
	\begin{assumption}
		\em \lab{ass3}
		There exists {\em at least} one $k \in \bar r$ such that ${\mathbb C}_{k,n_a+j}(t)=0,\;j \in \bar n_b.$
	\end{assumption}
	
	\begpro \em
	\label{pro3}
	Consider the LTV descriptor system \eqref{sys} and represent the system in the Standard Canonical form \eqref{weisys}, with the matrix ${\mathbb C}(t)$ verifying {Assumption \ref{ass3}}. Then, there exists a GPEBO of the form \eqref{hatza} such that \eqref{liminfa} holds provided an excitation condition stated in {Assumption \ref{ass5}} below is satisfied.
	\endpro
	
	\begproo
	First, notice that {Assumption \ref{ass4}} ensures the $k$-th output of the system is of the form\footnote{Clearly, if there are more outputs satisfying \eqref{yk} we can pile them up to create a regressor {\em matrix}---enhancing the possibility of satisfying the IE assumption imposed below.}
	\begequ
	\lab{yk}
	y_k=\begmat{{\mathbb C}_{k,1}(t) & {\mathbb C}_{k,2}(t) & \cdots & {\mathbb C}_{k,n_a}(t)}z_a=:\calc_k^\top(t) z_a.
	\endequ
	That is, there exists an output signal that we can compute without the knowledge of the unavailable state $z_b$. As explained in the proof of {Proposition \ref{pro1}}, this is a key step to generate the LRE used for the parameter estimator. 
	
	Proceeding with the GPEBO design we propose the dynamic extension
	\begalis{
		\dot \xi_a&= A_a(t) \xi_a+ f_a(t)\\
		\dot \Phi_a&=A_a(t) \Phi_a,\;\Phi_a(0)=I_{n_a},
	}
	and considering the dynamics \eqref{weisysa}, we can derive the identity
	$$
	z_a=\xi_a-\Phi_a \theta_a,
	$$
	with $\theta_a \in \rea^{n_a}$ and {\em unknown} vector. To obtain the regression equation we invoke \eqref{yk} and the unknown part of the state $z_a$ can be estimated via $\hat z_a=\xi_a-\Phi_a \hat \theta_a$, where the estimate $\hat \theta_a$ is generated using the LRE
	$$
	\caly_a=\psi^\top_a \theta_a,
	$$
	where we defined the signals
	\begalis{
		\caly_a &:=y_k-\calc_k^\top(t) \xi_a \\
		\psi_a &:=-\calc_k(t) \Phi_a(t).
	}
	A parameter estimate converging {\em exponentially fast} to its true value can be obtained with a DREM-based algorithm provided the following excitation assumption is statisfied.
	
	\begass \em
	\lab{ass5}
	The regressor vector $\psi_a(t)$ is IE.
	\endass
	\endproo
	
	\begrem
	\lab{rem2}
	As indicated in {\bf D3} above, for LTI systems the property of IE of the regressor constructed in GPEBO is determined by the observability property of the associated system. In the case above it is tantamount to the observability of the pair $(A_a,\calc_a)$. In that case, it is possible to design a classical Luenberger observer for the state $z_a$ measuring the output $y_k$ given in \eqref{yk}. For the LTV case,  we need to invoke Kalman-Bucy observers and impose the very strong {\em uniform complete observability} assumption---see \cite{WANORTBOB} for a detailed discussion of this point. As shown in \cite{WANORTBOB} the interest of implementing a GPEBO is that we can relax this strong assumption with the only requirement of {\em plain observability}---which is, of course, {\em necessary} to design (an on- or off-line) globally convergent observer. Moreover, as shown above GPEBO provides the possibility to deal with parameter uncertainty, a case in which the design of a Kalman-Bucy observer is far from clear. 
	\endrem 
	\subsection{GPEBO-based observer for (part of) $z_b$}
	\lab{subsec43}
	Similarly to the developments of Subsection \ref{subsec32} we restrict ourselves in this subsection to the case $n_b=3$ and treat the case when there are no uncertain parameters. Hence, we consider the second order dynamics \eqref{zbsys3}. As shown in Subsection \ref{subsec32}, under {Assumption \ref{ass3}}, the dynamics of the states $z_{b,1}$ and $z_{b,2}$ can be written in the form \eqref{dotzw}, and we will be interested in estimating {\em only} these two states.
	
	To complete the description of the system we need to incorporate an output signal. To clarify the scenario, let us write explicitly the output vector \eqref{weic} for $n_b=3$
	$$
	y=C_a(t)z_a+\begmat{C^b_{1,1}(t) \\ C^b_{2,1}(t) \\  \vdots \\ C^b_{r,1}}z_{b,1}+\begmat{C^b_{1,2}(t) \\ C^b_{2,2}(t) \\  \vdots \\ C^b_{r,2}}z_{b,2} +\begmat{C^b_{1,3}(t) \\ C^b_{2,3}(t) \\  \vdots \\ C^b_{r,3}}z_{b,3}.
	$$
	We make the following assumption pertaining to the matrix ${C}_b(t)$. 
	
	\begin{assumption}\em
		\lab{ass6}
		There exists {\em at least} one $\ell \in \bar r$ such that $C^b_{\ell,1}(t) \neq 0$, $C^b_{\ell,2}(t) \neq 0$ and $C^b_{\ell,3}(t)= 0.$
	\end{assumption}
	
	{Assumption \ref{ass6}} is required for three reasons: (i) the availability of an ``adequate" output signal; (ii) the verification of a ``relative degree"-like condition needed for the observer design;\footnote{We use the name ``relative degree" of the LTV system \eqref{dotzw}, \eqref{yw} with a slight abuse of notation---extrapolating it from the LTI case. See \cite[Definition 4.10]{RUGbook} for a rigorous definition of the term in the LTV case.}  and (iii) to ensure stability of the observer designed below. \\
	
	\noindent [Reason (i)] To explain the first reason define the vector
	$$
	\begmat{C^b_{\ell,1}(t) \\ C^b_{\ell,2}(t)}=:C_w(t) = \begmat{C_{w,1}(t) \\ C_{w,2}(t)}  
	$$
	and the signal 
	\begequ
	\lab{yw}
	y_w:=C_w^\top(t) z_w.
	\endequ
	Notice that, due to the condition $C^b_{\ell,3}(t)= 0$ we have that
	$$
	y_w=y_\ell-\sum_{i=1}^{n_a}C^a_{\ell,i}z_{a,i},
	$$
	with the unmeasurable term $z^b_{\ell,3}$ {\em absent}. We assume in the sequel that the state $z_a$ is estimated (exponentially fast) via GPEBO as described in Subsection \ref{subsec42}. Thus, replacing in a certainty-equivalent way, $z_a$ by its (easily obtained) estimated value and neglecting the exponentially decaying error term $\hat z_a(t)-z_a(t)$, we can assume that $y_w$ is {\em measurable}. This is the output that will be associated to the system \eqref{dotzw}.\\
	
	\noindent [Reason (ii)] Regarding the ``relative degree" issue (ii) notice that $C_w^\top(t)B_w = C^b_{\ell,2} \neq 0$. Consequently, the ``relative degree" of the triplet $(C_w(t),A_w(t),B_w(t))$ is {\em one}. In the presentation below we restrict ourselves to this case. Although, it is possible to treat also the case of ``relative degree" larger than one, the computations are extremely involved, leading to observer designs of little practical interest.\\
	
	\noindent [Reason (iii)] Finally, the stability-related issue (iii) is explained in the proof of the proposition below.\\
	
	The following assumption is required in the proof of the proposition below.
	
	\begass \em
	\lab{ass7}
	There exists a vector $L_w(t) \in \rea^2$ such that the two-dimensional LTV system
	$$
	\dot \varphi=\begmat{ \varphi_{11} & \varphi_{12} \\ \varphi_{21}  & \varphi_{22} } \varphi,
	$$
	where $\varphi_{11}=L_{w,1}(t)C_{w,1}(t)$, $\varphi_{12}=L_{w,1}(t)C_{w,2}(t)-a_{1}(t)$, $\varphi_{21}=L_{w,2}(t)C_{w,1}(t)+{\dot C_{w,1}(t) \over C_{w,1}(t)}$,
	$\varphi_{22}=L_{w,2}(t)C_{w,2}(t)+{1 \over C_{w,2}(t)}[\dot C_{w,2}(t)+ C_{w,1}(t) a_1(t)]$
	is {\em exponentially stable}.
	\endass
	
	\begpro\em
	\lab{pro4}
	Consider the system \eqref{dotzw}, \eqref{yw} verifying {Assumption \ref{ass5}} and with $\dot C_{w,2}(t)$ {\em measurable}. Define the second order dynamic observer
	\begsubequ
	\lab{sysobs}
	\begali{
		\lab{sysobsr}
		\nonumber
		\dot r &= M_w(t) r +[M_w(t) G_w(t)+L_w(t)-\dot G_w(t)]y_w\\
		&+[I_2-G_w(t) {C_w^\top}(t)]d_w(t)\\
		\hat z_w &=r+G_w(t)y_w,
		\lab{sysobsw}
	}
	\endsubequ
	with $L_w(t)$ a {\em free} vector an
	\begalis{
		M_w(t)&:=[I_2-G_w(t)C^\top_w(t)] A_w(t) -L_w(t) C_w^\top(t)\\
		& -G_w(t)\dot C^\top_w(t)\\
		G_w(t)&:=\begmat{0\\{1 \over C_{w,2}(t)}}.
	}
	
	For all systems initial conditions,
	\begequ
	\lab{conconw}
	\liminf |\hat z_w(t)-z_w(t)|=0,
	\endequ
	exponentially fast, provided $L_w(t)$ satisfies {Assumption} \ref{ass7}.
	\endpro
	
	\begin{proof}
		To establish the proof we derive the dynamics of the observation error $\tilde z_w:=\hat z_w-z_w$. Hence, we  compute
		\begin{footnotesize}
			\begalis{
				&\dot {\tilde z}_w = \dot r+ \dot G_w(t)y_w+G_w(t) \dot y_w - \dot {z}_w\\
				=& M_w(t)\hat z_w+[I_2-G_w(t) {C_w^\top}(t)]d_w(t)+ L_w(t)y_w\\
				&+G_w(t)\dot y_w- [A_w(t) z_w +d_w(t)+B_w \zeta]\\
				=&[A_w(t)-L_w(t) {C_w^\top}(t)-G_w(t)\dot C^\top_w(t)-G_w(t) C^\top_w(t) A_w(t)]\hat z_w\\
				& +[I_2-G_w(t) {C_w^\top}(t)]d_w(t)+L_w(t) y_w\\
				& + G_w(t)\{[\dot C^\top_w(t)+ C^\top_w(t) A_w(t)] z_w + C_w^\top(t)[d_w(t)+B_w \zeta]\}\\
				&- [A_w(t) z_w +d_w(t)+B_w \zeta]\\
				=& \{[I_2-G_w(t)C^\top_w(t)] A_w(t) -L_w(t) C_w^\top(t) -G_w(t)\dot C^\top_w(t)\} \tilde z_w\\
				=& M_w(t) \tilde z_w,
			}
		\end{footnotesize}
		
		where we used the fact that
		$$
		[G_w(t)\ C^\top_w(t)-I_2]B_w=0,
		$$
		to get the last identity.
		
		Now, after some lengthy but straightforward calculations, it is possible to prove that 
		$$
		M_w(t)=\begmat{M_{11} &  M_{12} \\ M_{21}  & M_{22} }.  
		$$
		where $M_{11}=L_{w,1}(t)C_{w,1}(t)$, $M_{12}=L_{w,1}(t)C_{w,2}(t)-a_{1}(t)$, $M_{21}=L_{w,2}(t)C_{w,1}(t)+{\dot C_{w,1}(t) \over C_{w,1}(t)}$, $M_{22}=L_{w,2}(t)C_{w,2}(t)+{1 \over C_{w,2}(t)}[\dot C_{w,2}(t)+ C_{w,1}(t) a_1(t)]$.
		
		At this point we make the observation that, if the coefficient $C_{w,1}(t) \equiv 0$, then the matrix $M_w(t)$ cannot satisfy {Assumption} \ref{ass7}. This situation is ruled by {Assumption} \ref{ass6}. The proof is completed invoking {Assumption} \ref{ass7}.
	\end{proof}
	\begrem
	\lab{rem3}
	A sensible choice for $L_w(t)$ is given by
	$$
	L_{w}(t)=\begmat{{a_1(t) \over C_{w,2}(t)} \\ \\ -{\dot C_{w,1}(t) \over C^2_{w,1}(t)}}
	$$
	which {\em diagonalizes} the matrix $M_w(t)$. Then, the stability test boils down to integrability of the resulting diagonal terms. 
	\endrem
	
	\begrem
	\lab{rem4}
	It is possible to prove that in the LTI case {Assumption} \ref{ass7} is satisfied with the choice
	$$
	L_w=a_{1} C_{w,1} \begmat{{C_{w,1} \over C_{w,2}} \\ 1},
	$$
	provided $ a_{1} {C_{w,1}\over C_{w,2}}<0.$ From \eqref{dotzw} it is clear that it is possible to ``change the sign" of the coefficient $a_1$ introducing a change of coordinates $z_w'=\begmat{1 & 0 \\ 0 & -1}z_w$. Hence, {Assumption} \ref{ass7} can {\em always} be satisfied in the LTI case. Actually, in view of Remark \ref{rem1} the design of the observer of $z_b$ for LTI systems can be significantly simplified.
	\endrem
	%
	\section{A Benchmark LTV Circuit Example} 
	\lab{sec5}
	%
	In \cite[Chapter 6]{BOB}---see also \cite[Section 7]{BOBCAM}---the circuit example depicted in Fig. \ref{fig2} is considered, with the capacitors, inductor and resistors {\em time-varying}. A state observer is designed invoking the systems completion procedure that, as explained in the Introduction, entails the calculation of signal derivatives. Here, we propose a GPEBO-based solution.
	
	\begin{figure}[H]
		\centering
		\includegraphics[width = 0.45\textwidth]{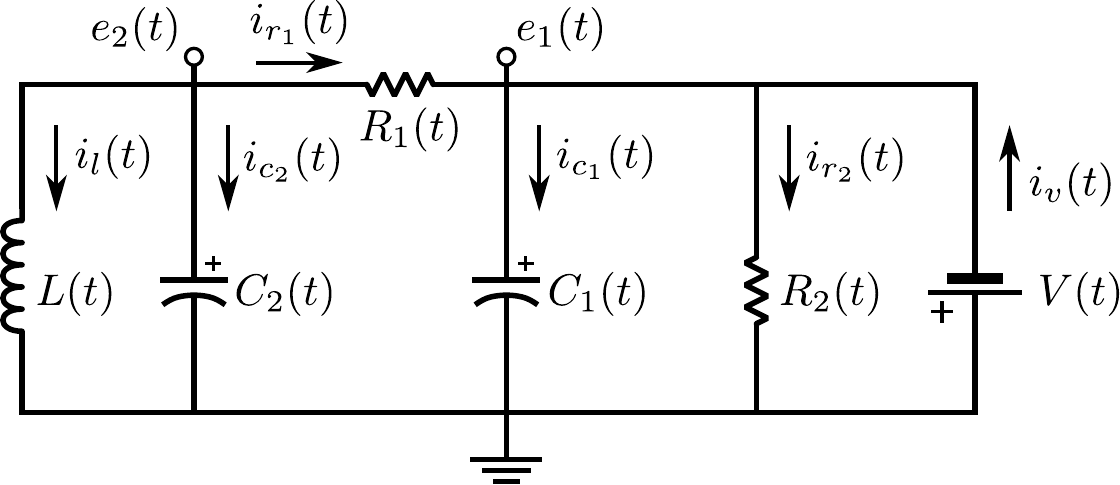}
		\caption{Circuit example from \cite[Chapter 6]{BOB}.} 
		\label{fig2}
	\end{figure}
	
	As shown in \cite{BOBCAM} the dynamic behavior of the circuit may be written in the form 
	\begalis{
		E(t) \dot x &= A(t)x + Bu.
	}
	where $x:=\col(e_1,e_2,i_l,i_{r_1},i_{r_2},i_v)$, $u:=V$ and
	\begin{align*}
		E(t) &= \begmat{
				{C_a(t) } & 0 & 0 & 0 & 0 & 0\\
				0 & {C_b(t) } & 0 & 0 & 0 & 0\\
				0 & 0 & {L(t)} & 0 & 0 & 0\\
				0 & 0 & 0 & 0  & 0 & 0\\
				0 & 0 & 0 & 0 & 0  & 0\\
				0 & 0 & 0 & 0 & 0 & 0}, B=\begmat{0 \\ 0 \\ 0 \\ 0 \\ 0 \\-1}\\
		A(t)&=\begmat{
			-\dot{C}_1(t)  & 0 & 0 & 1 & -1 & 1\\
			0 & -\dot{C}_2(t) & -1 & -1 & 0 & 0\\
			0 & 1 & -\dot{L}(t) & 0 & 0 & 0\\
			-1 & 1 & 0 & R_1(t)  & 0 & 0\\
			-1 & 0 & 0 & 0 & R_2(t)  & 0\\
			-1 & 0 & 0 & 0 & 0 & 0}. 	
	\end{align*}
	
	Similarly to \cite{BOBCAM} we consider that  the output matrix is of the form
	$$
	C=\begmat{0 & 0 & 1 & 0 & 0  & 1}.
	$$
	
	Partitioning $x=\col(x_a,x_b)$, with $x_a(t) \in \rea^3,x_b(t) \in \rea^3$, we can rewrite our system in the form\footnote{To avoid cluttering the notation, we omit in the sequel the time argument.}
	\begalis{
		\dot x_a&=A_{11} x_a+A_{12} x_b+B_1u\\
		0&=A_{21} x_a+A_{22} x_b+B_2u\\
		y&=C_{a} x_a+C_{b} x_b,
	}
	with 
	\begin{align*}
		A_{11} &:= \begmat{
			-\frac{\dot{C}_1}{C_1 } & 0 & 0\\
			0 & -\frac{\dot{C}_2}{C_2 } & -\frac{1}{C_2 }\\
			0 & \frac{1}{L} & -\frac{\dot{L}}{L}},\;
		A_{12}=\begmat{
			\frac{1}{C_1 } & -\frac{1}{C_1 } & \frac{1}{C_1 }\\
			-\frac{1}{C_2 } & 0 & 0\\
			0 & 0 & 0}, \\
		A_{21} &:= \begmat{
			-1 & 1 & 0\\
			-1 & 0 & 0\\
			-1 & 0 & 0},\;
		A_{22} := \begmat{
			R_1  & 0 & 0\\
			0 & R_2  & 0\\
			0 & 0 & 0},
	\end{align*}
	and $C_a=C_b:=\begmat{0 & 0 & 1}$. Similarly to the example of Subsection \ref{subsec21}, we assume that the capacitors and inductors are bounded away from zero, and rewrite the system as
	\begalis{
		\dot x_a&=A_{11} x_a+A_{12} x_b+B_1u\\
		\cala_{b} x_b &=\begmat{-B_2 u \\ y}-\begmat{A_{21} \\ C_a} x_a,
	}
	with 
	$$
	\cala_b:=\begmat{A_{22} \\ C_b}=\begmat{R_1 & 0 & 0\\ 0 & R_2 & 0 \\0 & 0 & 0\\0 & 0 & 1}.
	$$
	This matrix is clearly full rank and injective, hence it satisfies {Assumption 1}.  Consequently, to invoke Proposition \ref{pro1} for the design of the GPEBO, it only remains to verify the IE {Assumption 2} of the regressor $\psi$ defined in \eqref{regpsi}.
	
	For simulations we followed \cite[Chapter 6]{BOB} and considered the functions below for the circuit elements: 
	\begin{align*}
	C_1(t)&=3+\cos(t/3),\;C_2(t)=2-\cos(2t),\\
	L(t)&=2-\exp(-t),\\
	R_1(t) &= -(4+2\sin(t)),\;R_2(t)=-(2+\sin(t)),
	\end{align*}
	and the voltage input signal  $u(t)=4 \cos(2t) \sin(t/5)$. The initial conditions for the system were taken as $x_a(0)=\col(0, 1, 2)$ and ${x_b(0)=\col(0.25, 0, -0.25)}$ and for the estimators dynamic extension as $\xi_a(0)=\col(1, 0, 0)$. 
	
	Following \cite{ORTetaltac} the estimation was carried out with Kreisselmeier regression extension \cite[Section IV.B]{ORTetaltac}, the standard mixing step \cite[Section II.B]{ORTetaltac} and the classical gradient estimator \cite[Equation (10)]{ORTetaltac}. The filters were selected as $\calk(p)={\lambda \over p+\lambda}$ with $\lambda=\diag\{0.1,0.2,0.3\}$ and the gradient gain chosen as $\gamma=10^{10}$.
	
	The simulation results for the regressor $\psi$ are shown in Fig.~\ref{psi}, which clearly shows that it is IE. The behavior of the remaining signals of interest is depicted in Figs. \ref{fig3}-\ref{fig8}. To illustrate the effect of the adaptation gain $\gamma$ on the transient behavior, in Figs, \ref{fig6}-\ref{fig8} we show the result for two different values of this gain. As expected, increasing the gain accelerates the convergence.
	
	\begin{figure}[h]
		\center{\includegraphics[width=1\linewidth]{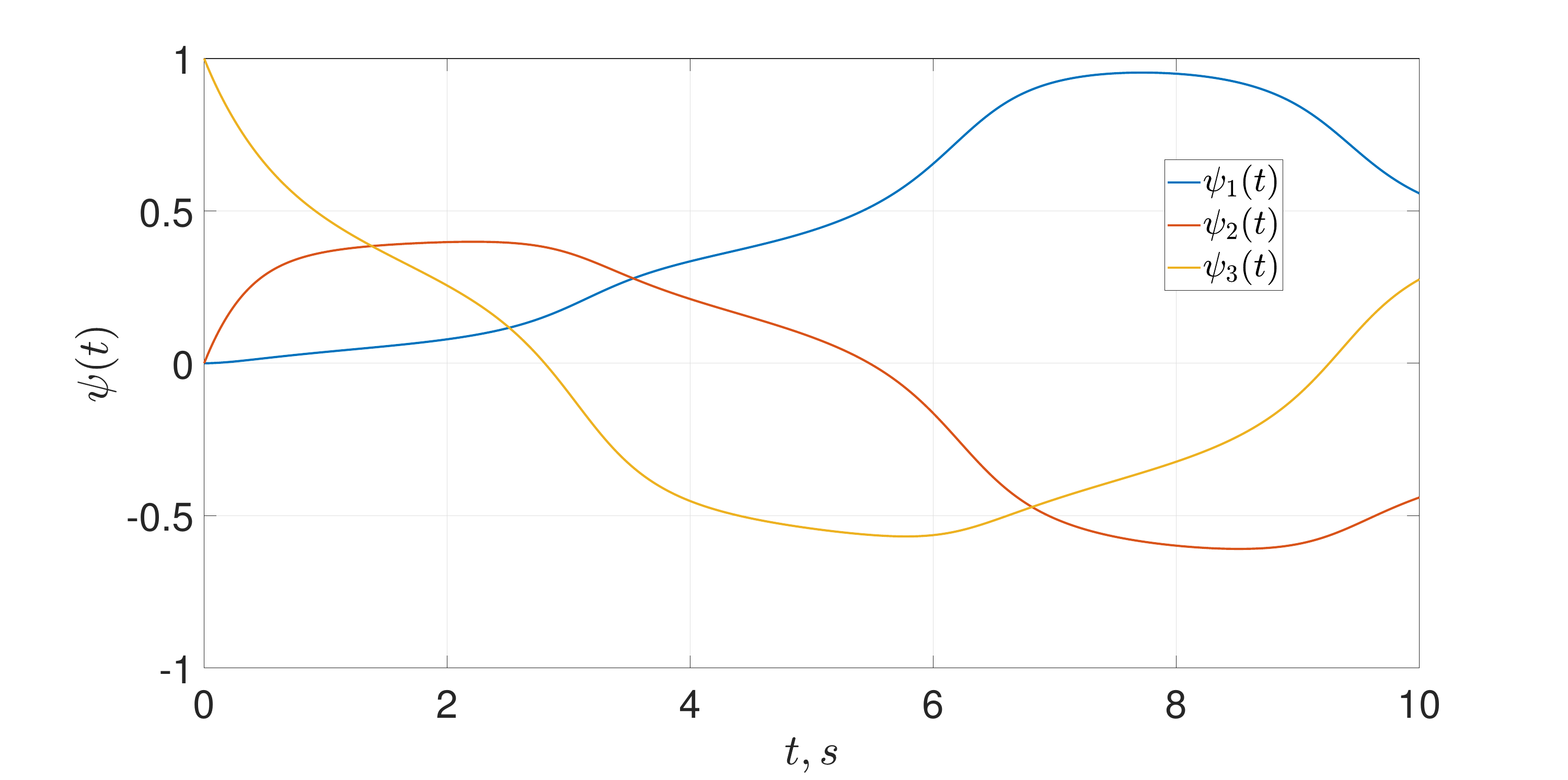}}
		\caption{Transients of the regressor $\psi$}
		\label{psi}
	\end{figure}
	
	\begin{figure}[h]
		\center{\includegraphics[width=1\linewidth]{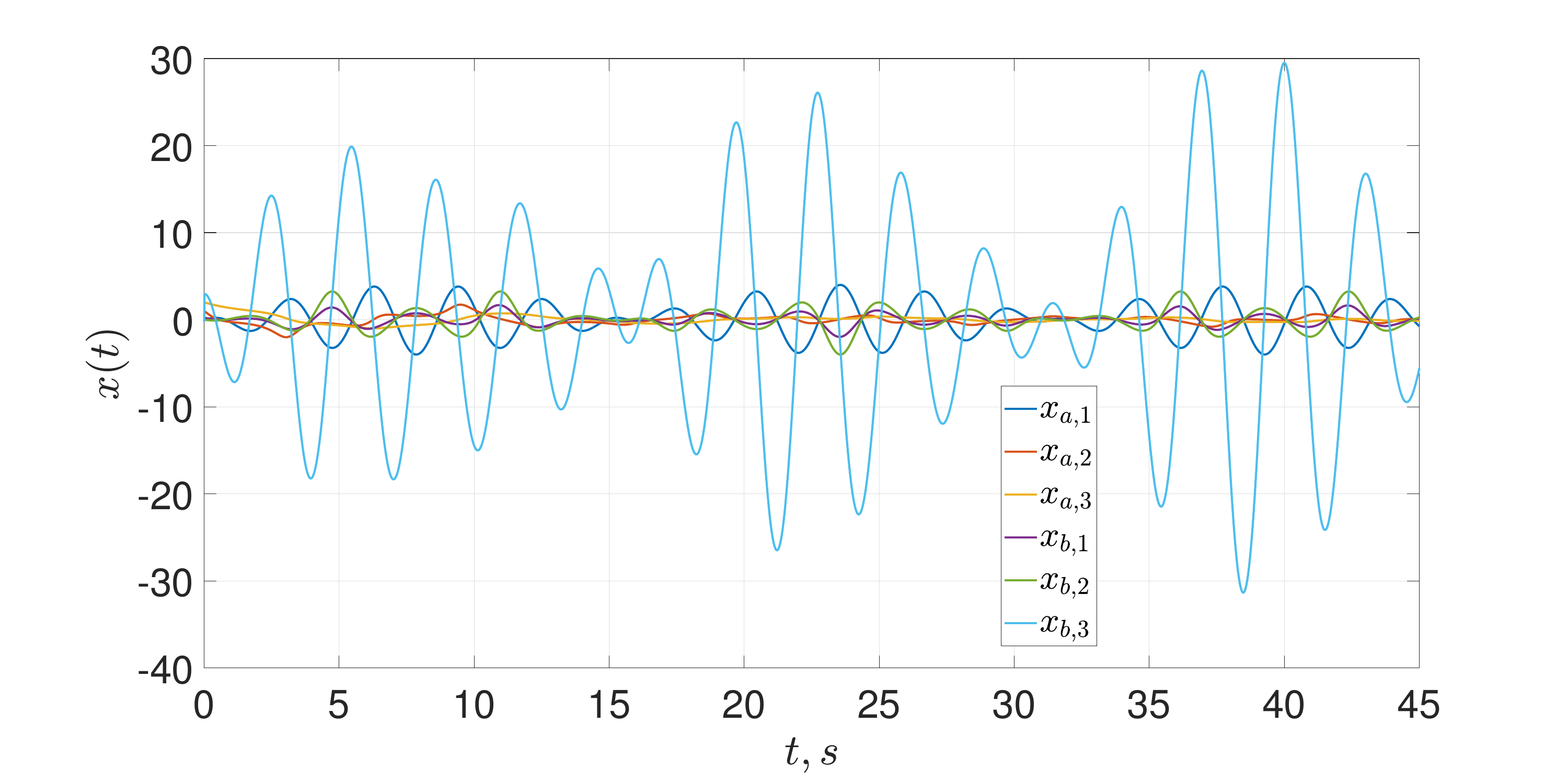}}
		\caption{Transients of the state vector}
		\label{fig3}
	\end{figure}
	
	\begin{figure}[H]
		\centering
		\includegraphics[width = 0.5\textwidth]{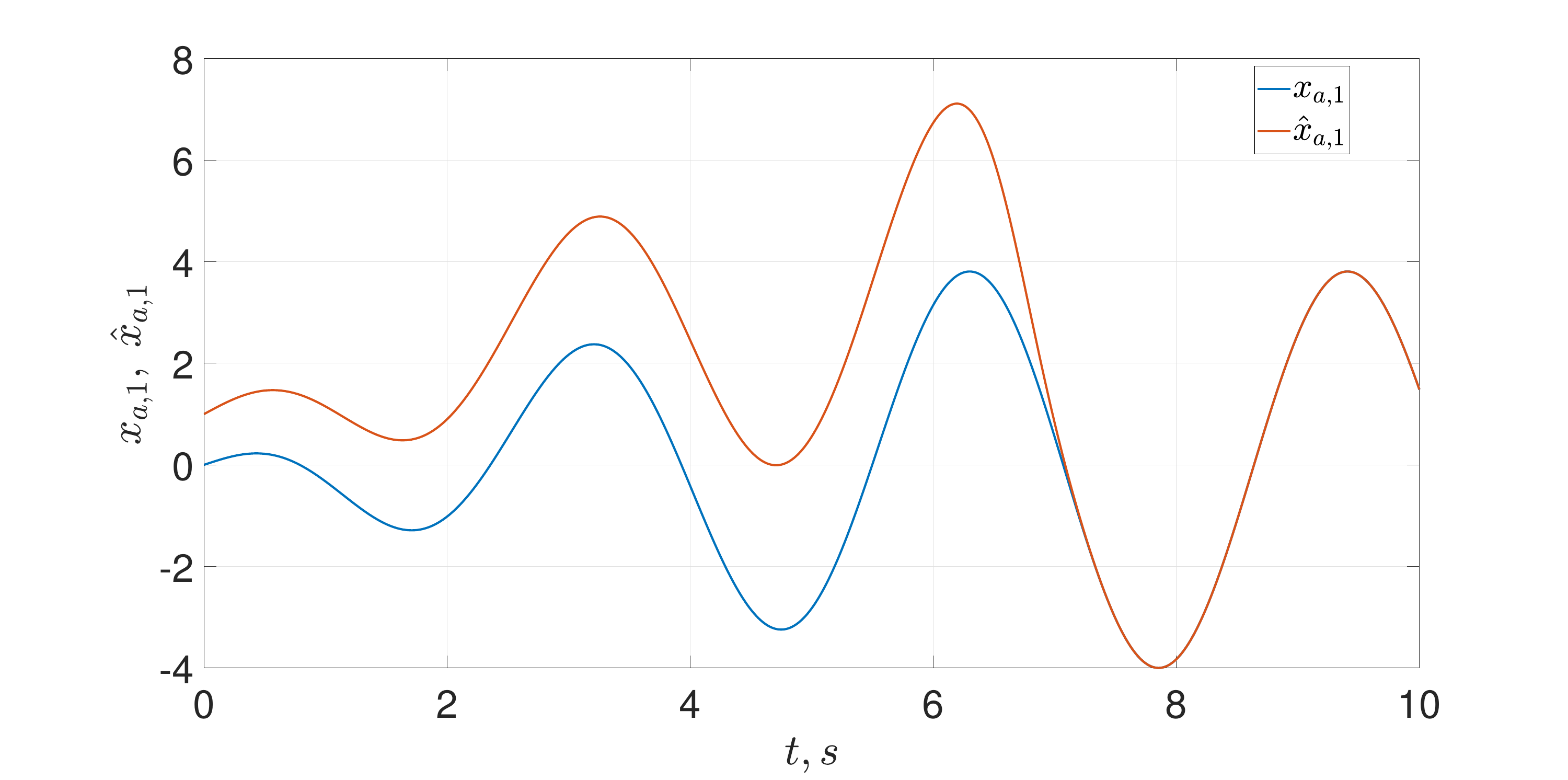}
		\centering
		\includegraphics[width = 0.5\textwidth]{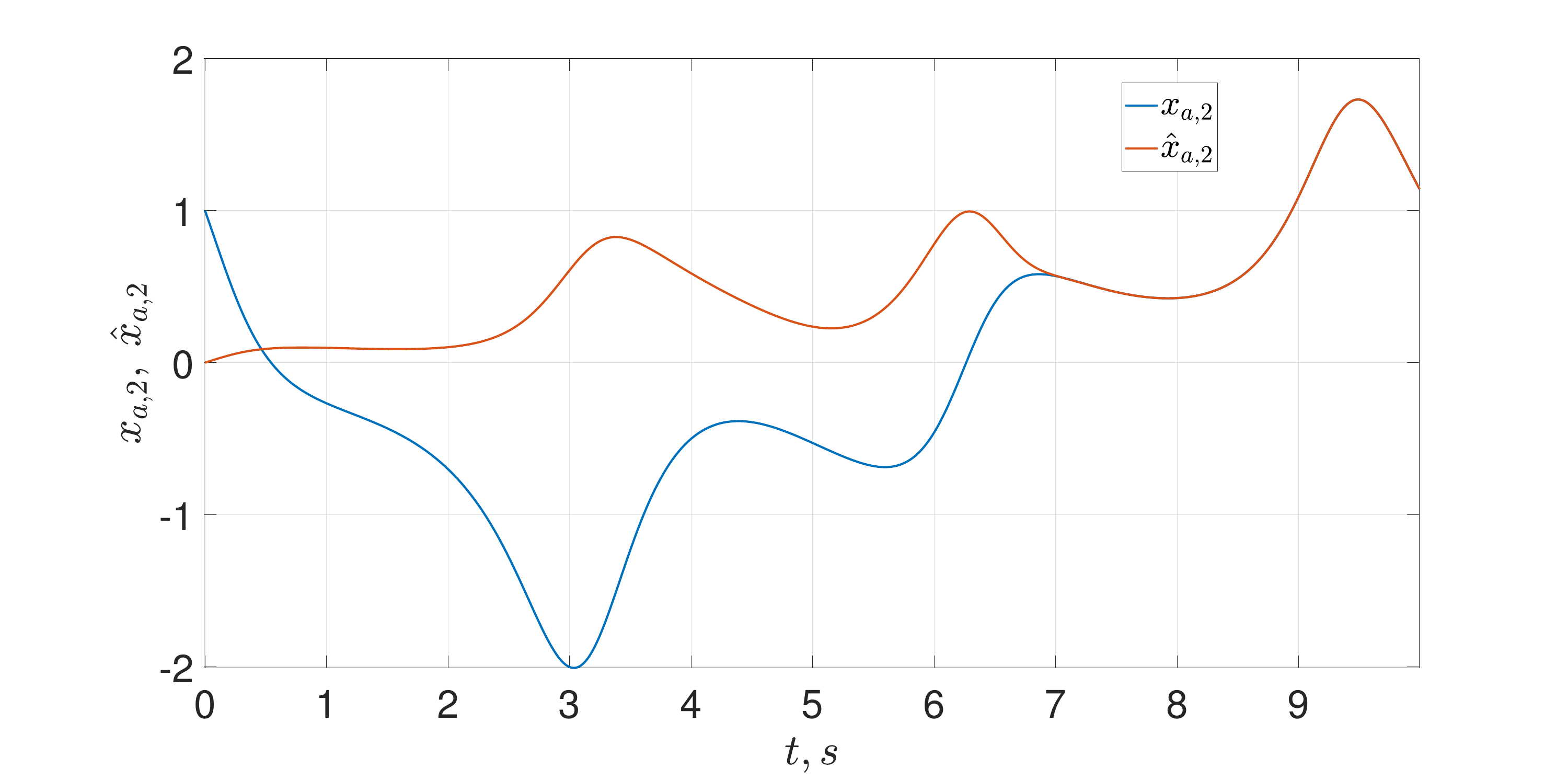}
		\centering
		\includegraphics[width = 0.5\textwidth]{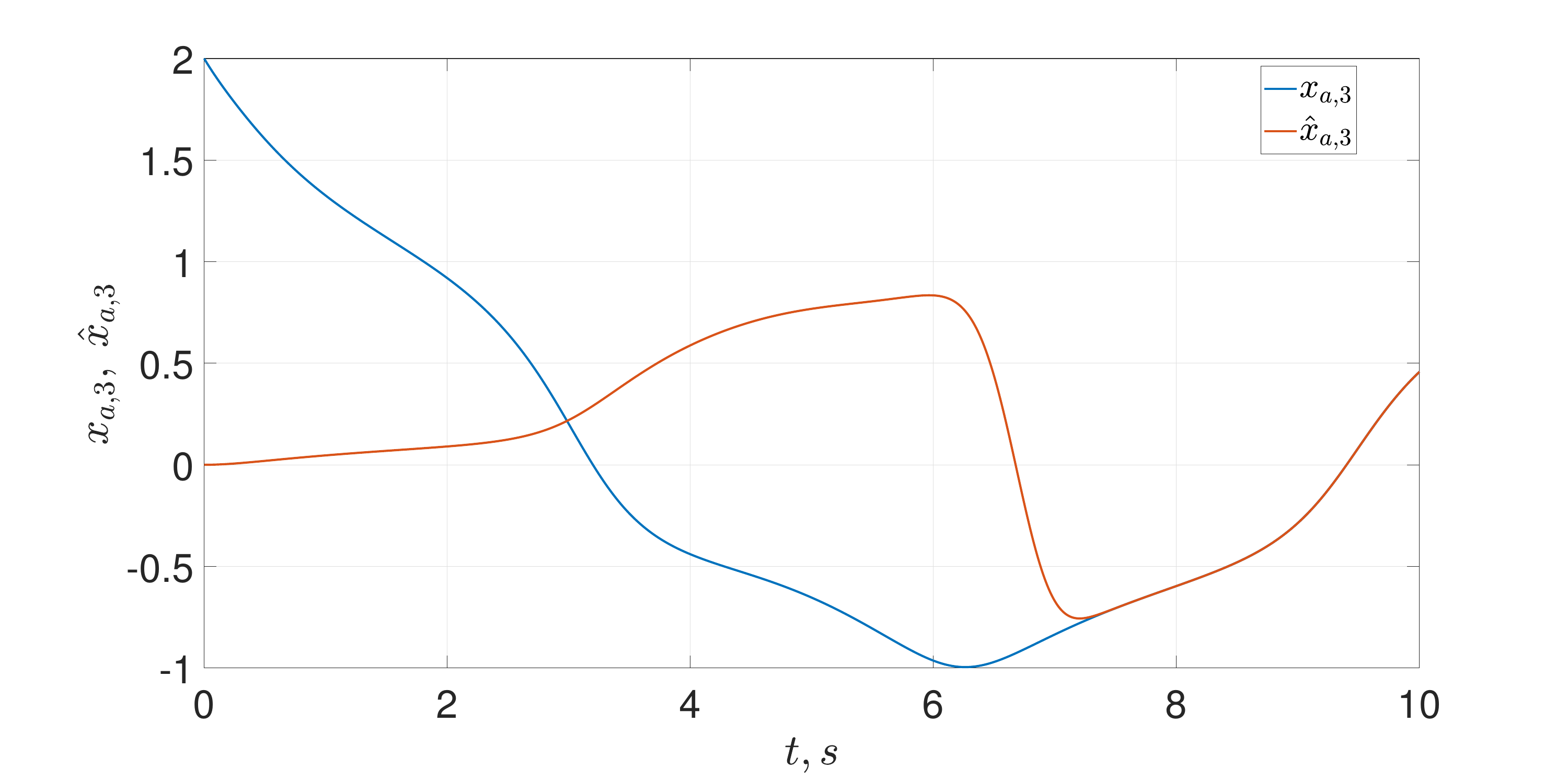}
		\caption{State $x_a$ and state estimation $\hat{x}_a$} 
		\label{fig4}
	\end{figure}
	
	\begin{figure}[H]
		\centering
		\includegraphics[width = 0.5\textwidth]{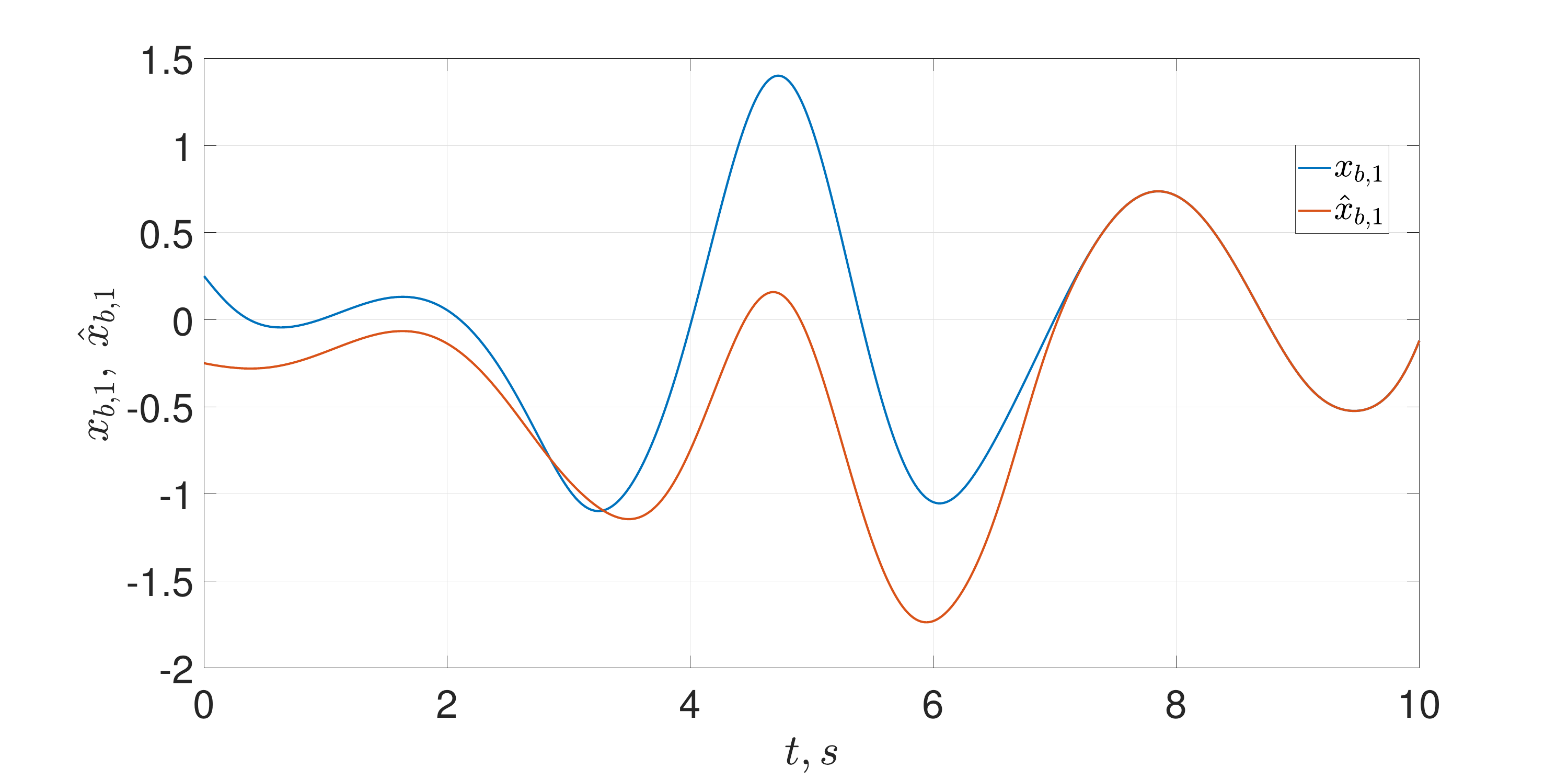}
		\centering
		\includegraphics[width = 0.5\textwidth]{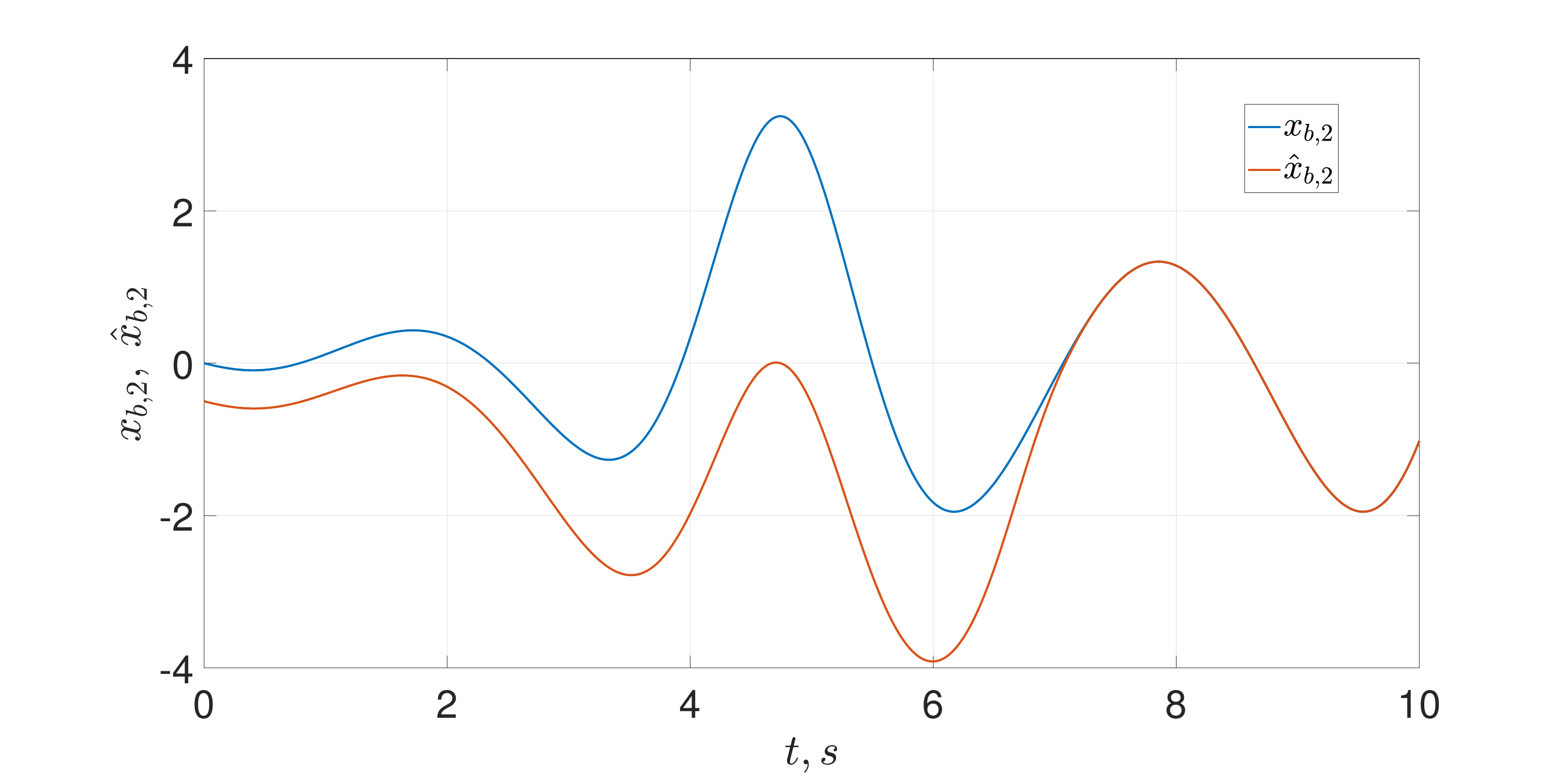}
		\centering
		\includegraphics[width = 0.5\textwidth]{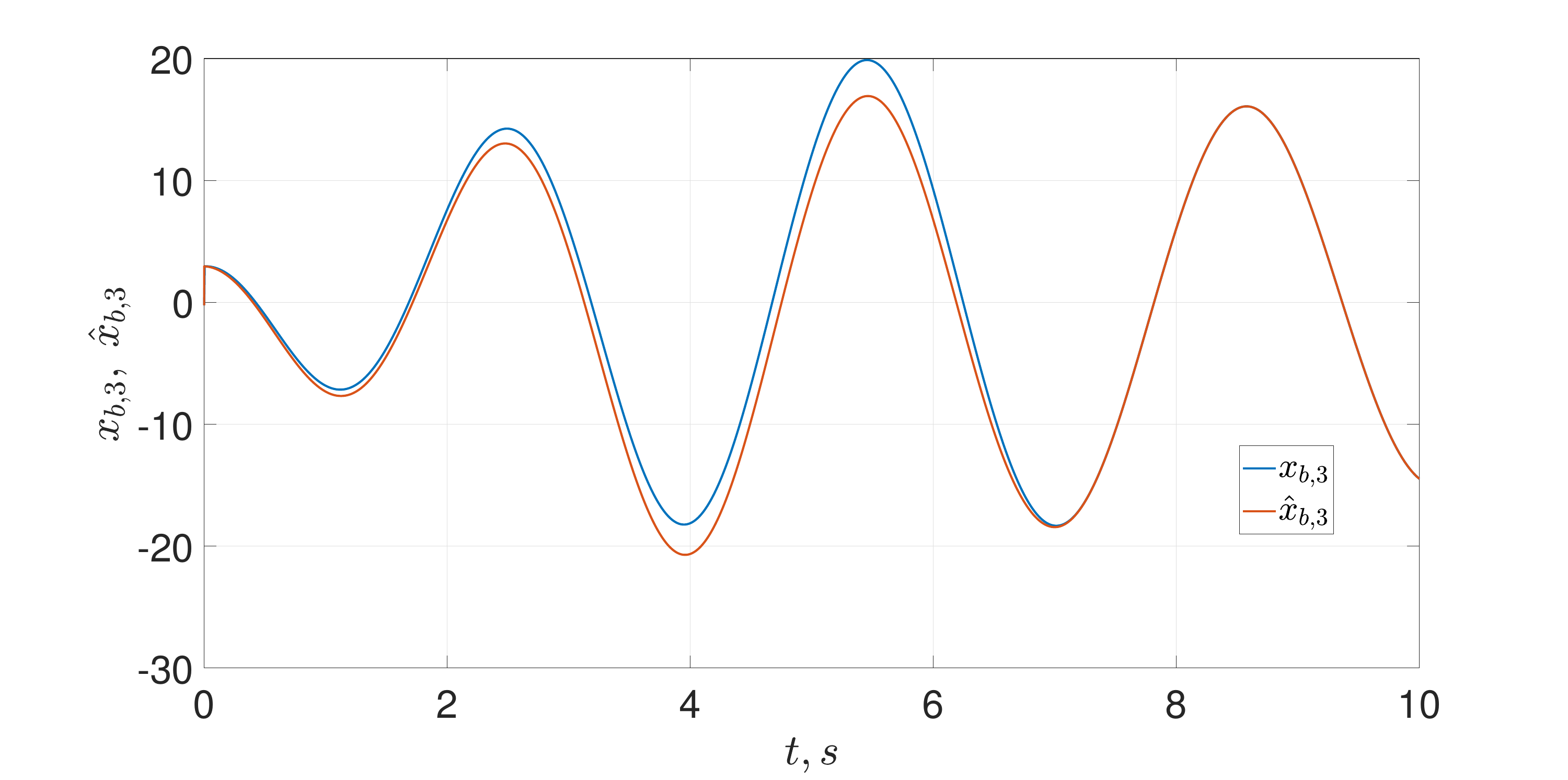}
		\caption{State $x_b$ and state estimation $\hat{x}_b$} 
		\label{fig5}
	\end{figure}
	
	\begin{figure}[H]
		\centering
		\includegraphics[width = 0.5\textwidth]{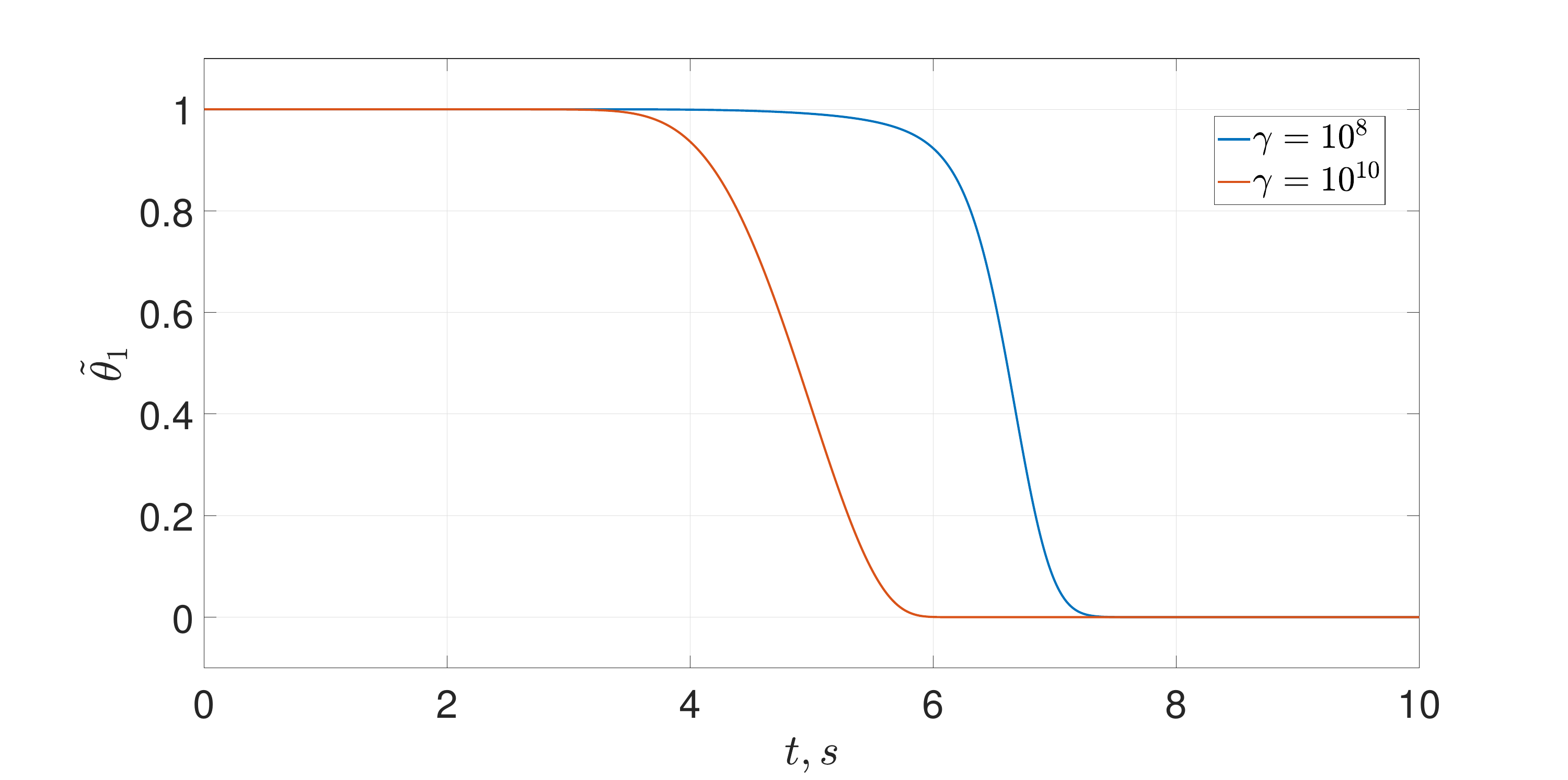}
		\centering
		\includegraphics[width = 0.5\textwidth]{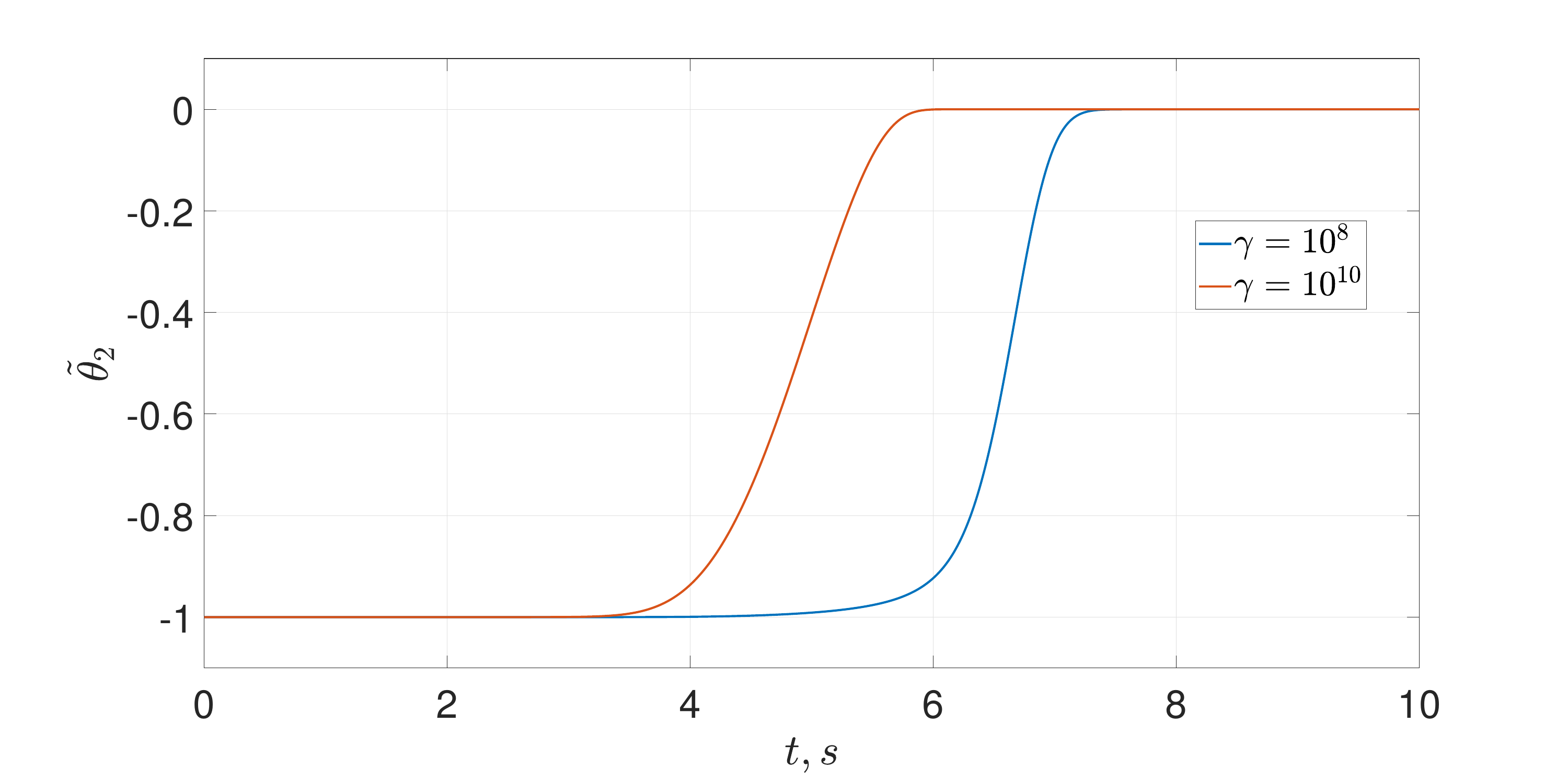}
		\centering
		\includegraphics[width = 0.5\textwidth]{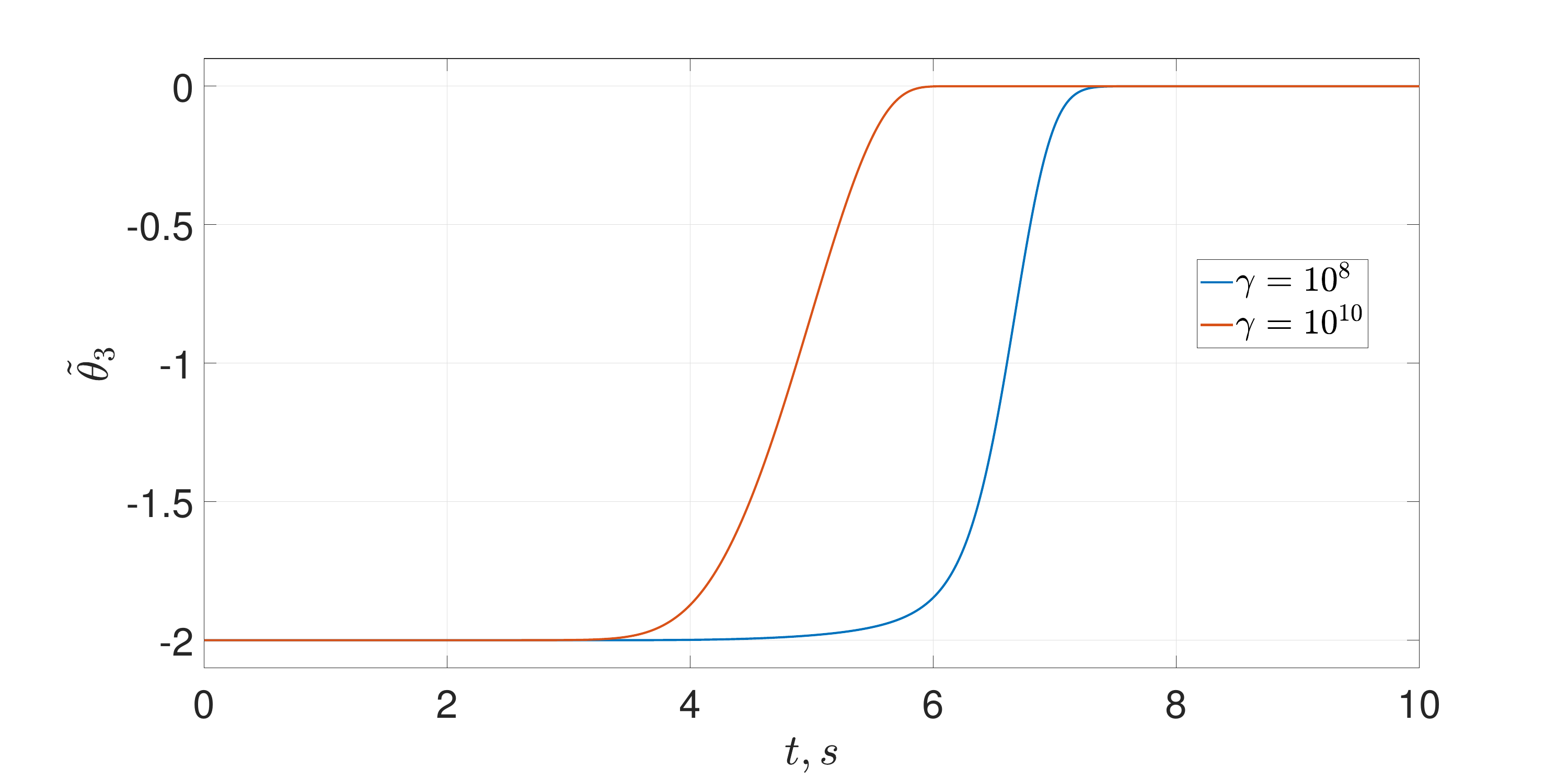}
		\caption{Initial conditions estimation error $\tilde \theta~:=~\hat{\theta}-\theta$, where $\theta~:=~\xi_a(0)-x_a(0)$,  for two gradient adaptation gains.} 
		\label{fig6}
	\end{figure}
	
	\begin{figure}[H]
		\centering
		\includegraphics[width = 0.5\textwidth]{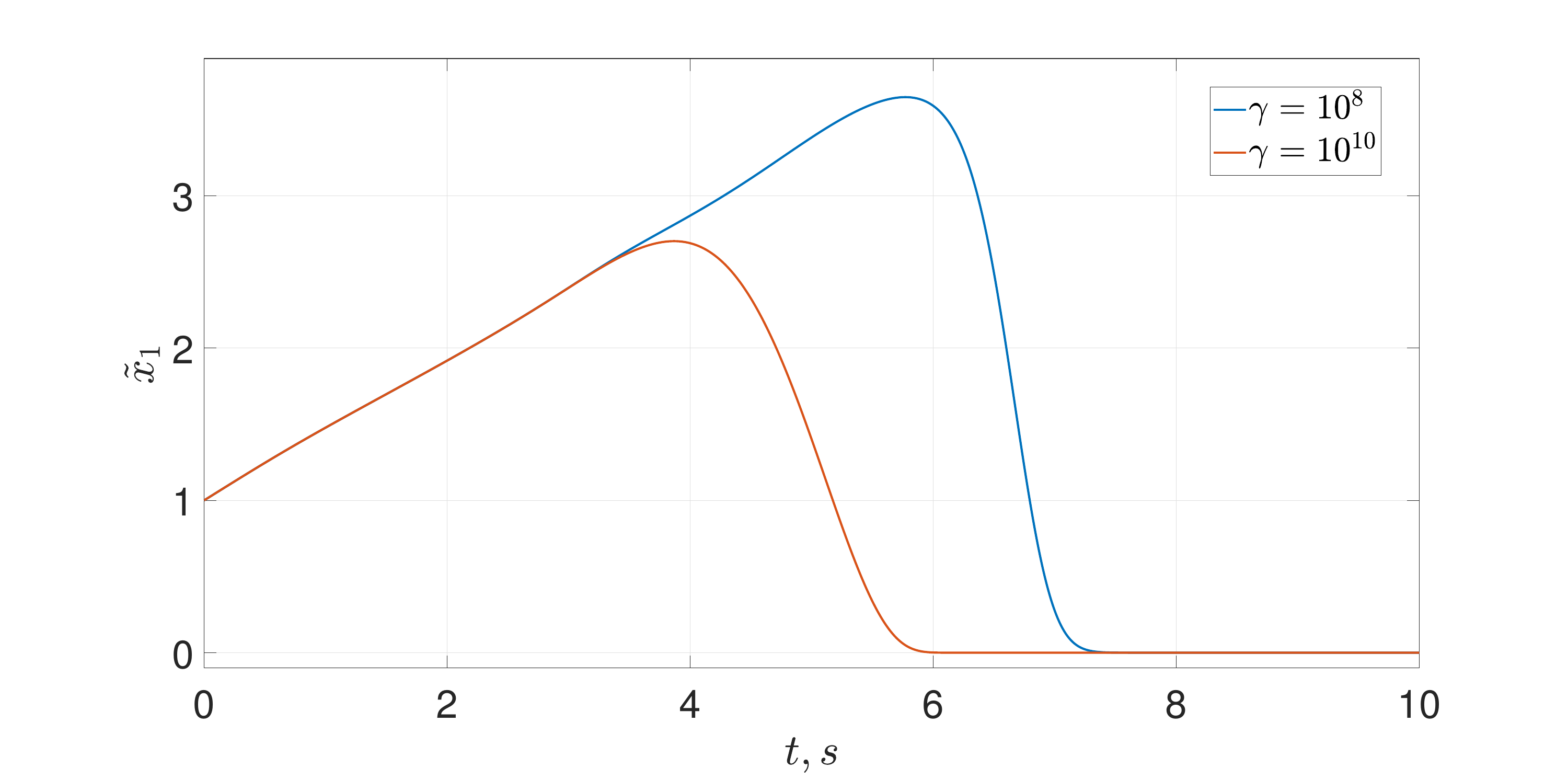}
		\centering
		\includegraphics[width = 0.5\textwidth]{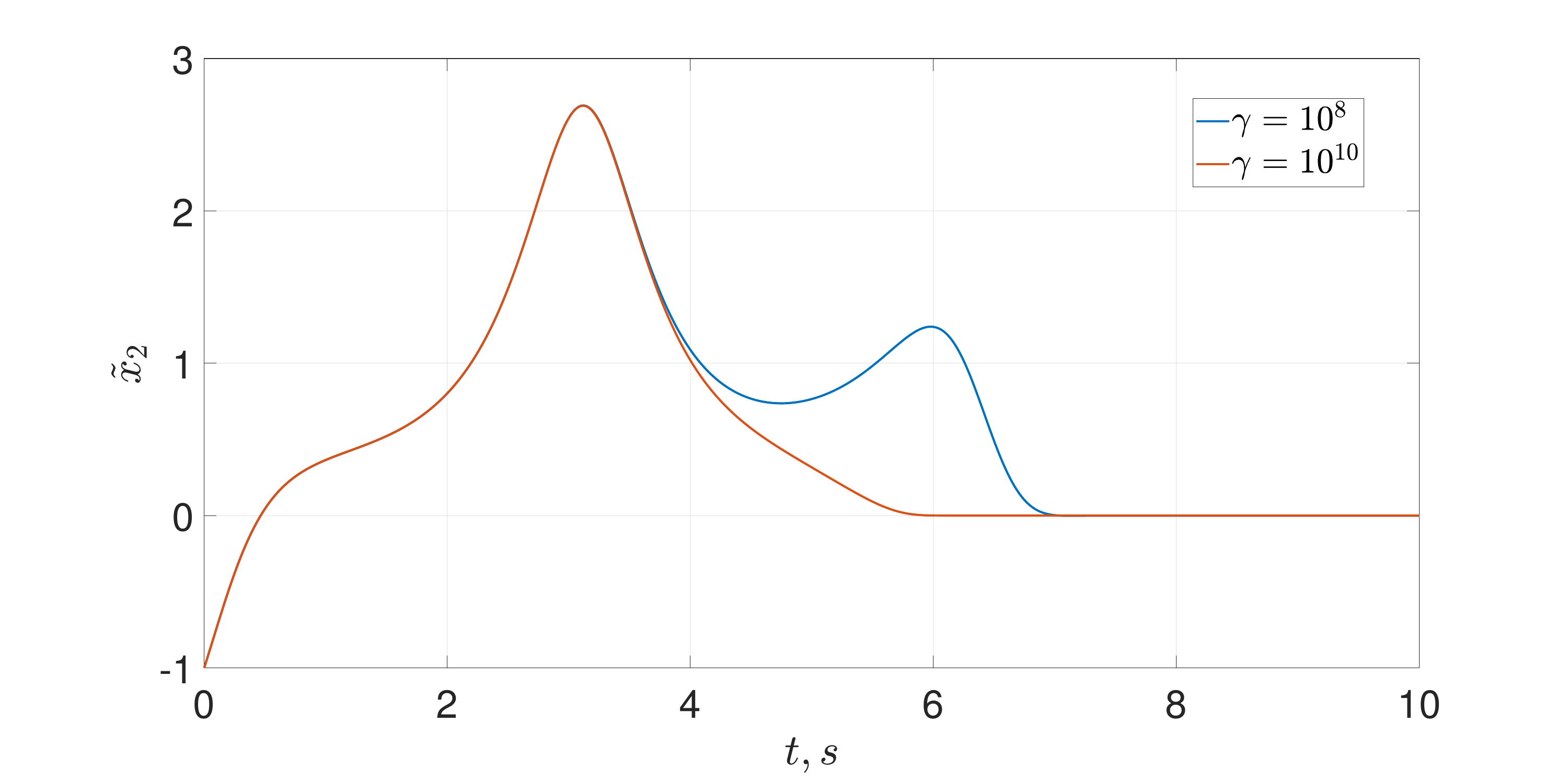}
		\centering
		\includegraphics[width = 0.5\textwidth]{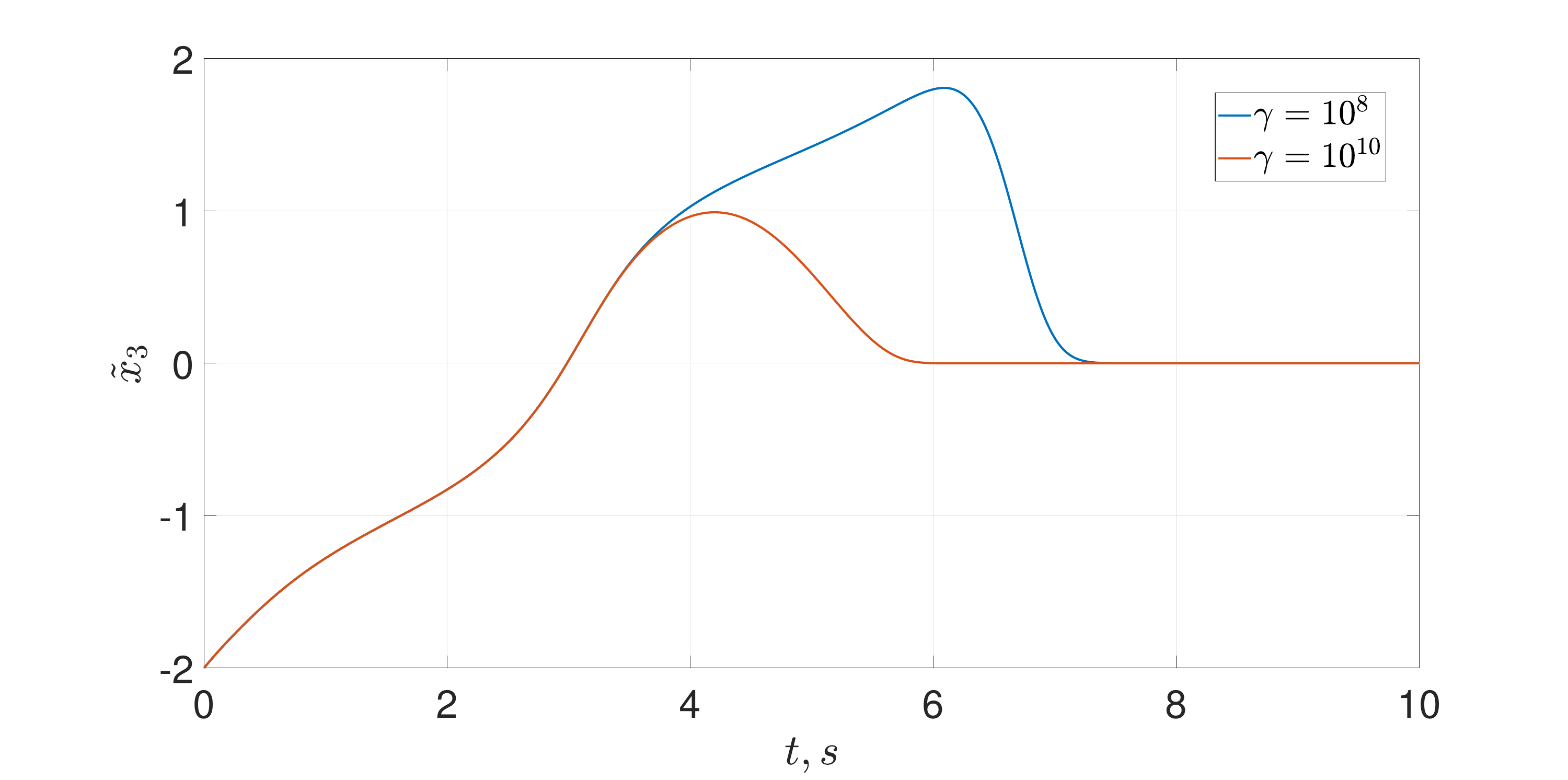}
		\caption{State estimation error $\tilde x_{a} :=\hat{x}_{a}-{x}_{a}$,  for two gradient adaptation gains.} 
		\label{fig7}
	\end{figure}
	
	\begin{figure}[H]
		\centering
		\includegraphics[width = 0.5\textwidth]{figures/err_x3}
		\centering
		\includegraphics[width = 0.5\textwidth]{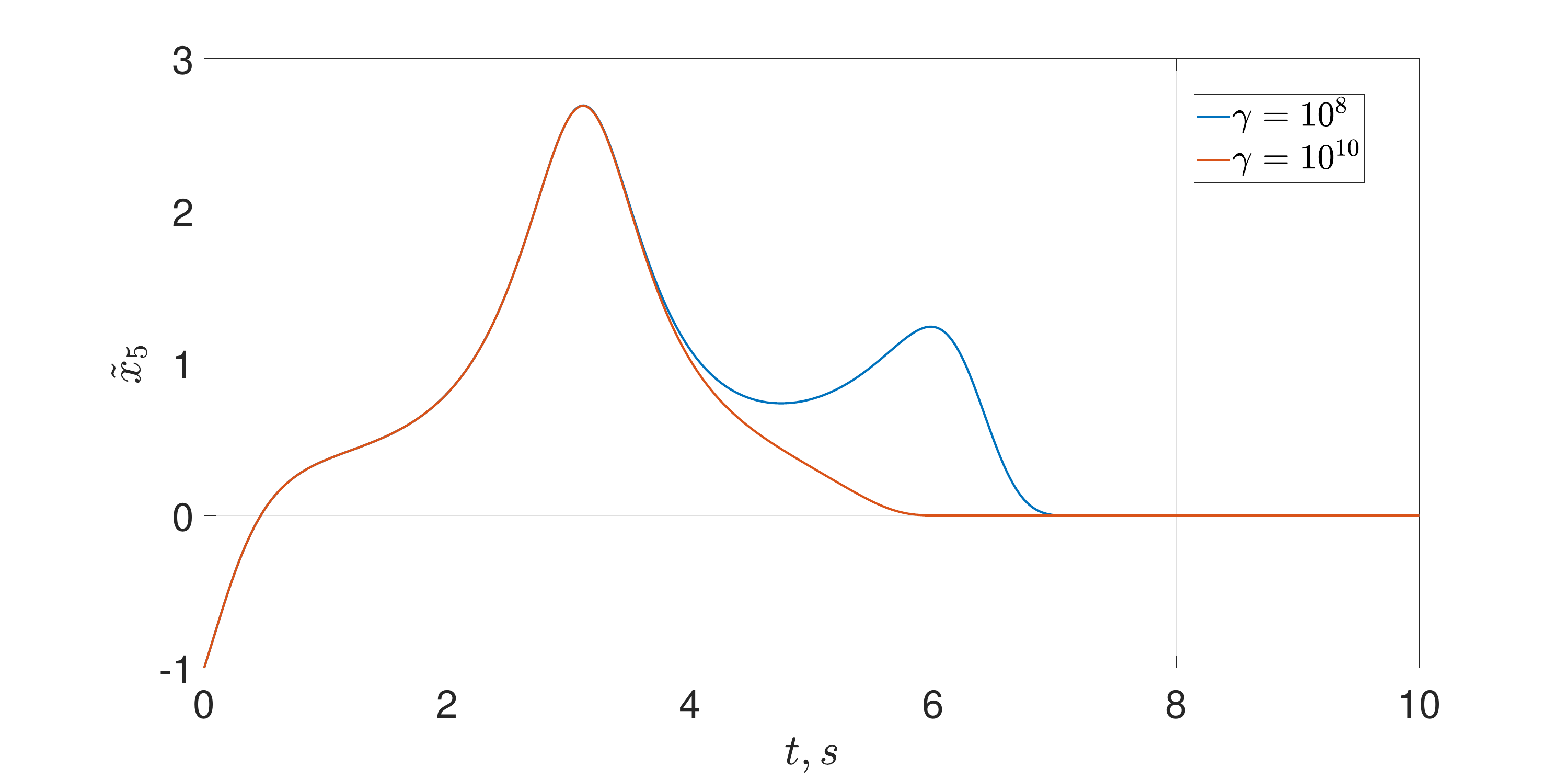}
		\centering
		\includegraphics[width = 0.5\textwidth]{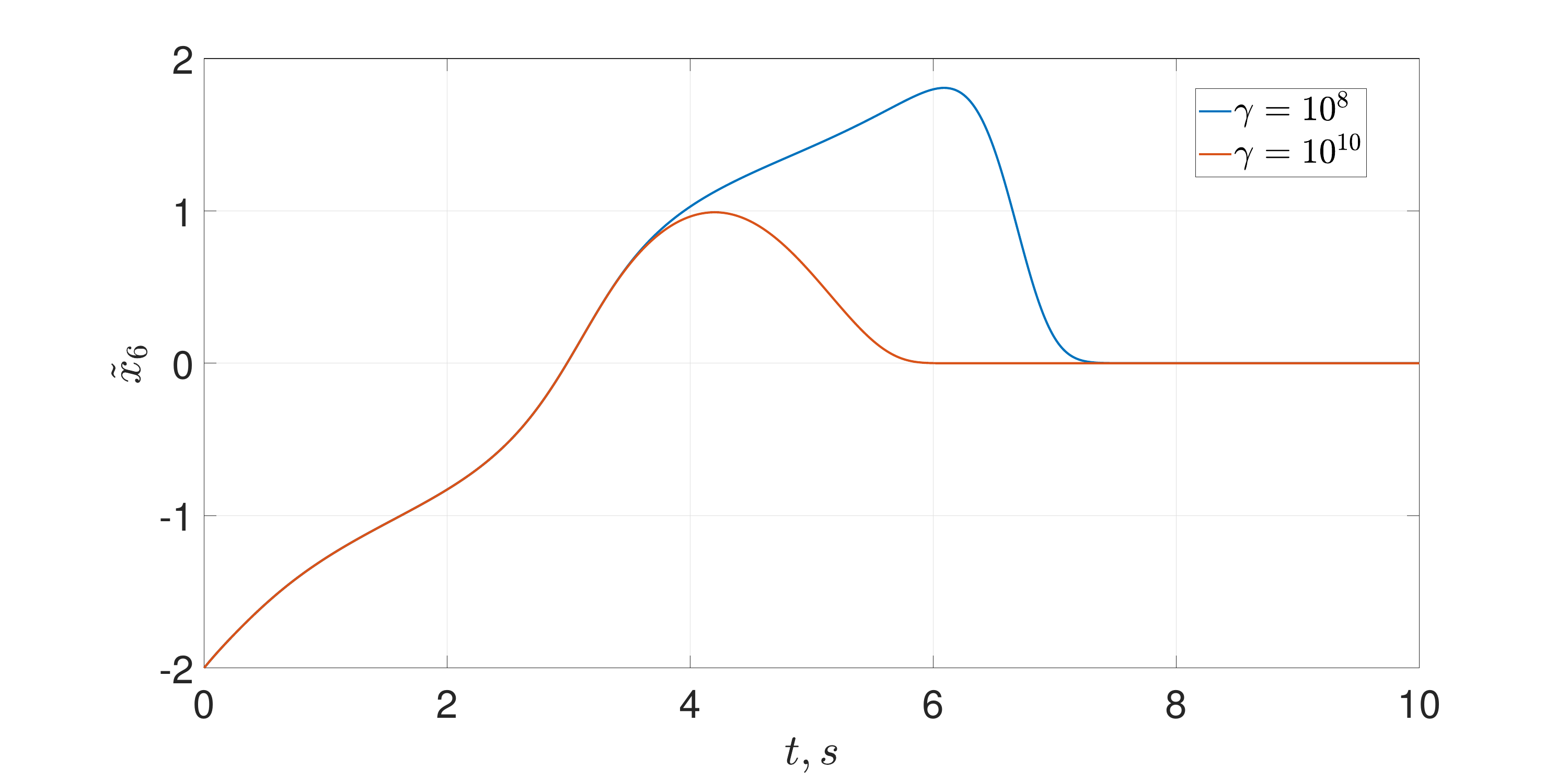}
		\caption{State estimation error $\tilde x_{b} :=\hat{x}_{b}-{x}_{b}$,  for two gradient adaptation gains.} 
		\label{fig8}
	\end{figure}
	%
	\section{Concluding Remarks and Future Research} 
	\lab{sec6}
	%
	We have presented in this paper a first attempt to develop a comprehensive theory for the design of observers for general LTV descriptor systems. A consequence  of our intention to keep the analysis as general as possible is that the final results are a little bit cryptic. Certainly, proceeding from this general framework, we can address particular cases and achieve some practically workable designs. 
	
	We have identified several critical aspects of the problem. For instance, the highly relevant role played by the strangeness index---see Subsection \ref{subsec41}. Intensive research has been carried out to develop procedures to render the system strangeness-free, see \cite{KUNMEHbook,MEHUNG} for some results along these lines. Our current work is aimed at applying these techniques for the problem of observer design and we expect to be able to report some results soon. We have also underscored the differences between the LTV and the LTI case, which are quite significant.
	
	Another research line that we are pursuing is to develop the theory for particular classes of LTV descriptor systems, with special emphasis on practical examples. In this respect a class of systems of great relevance are the ones described by port-Hamiltonian models.  

\appendices

\section{Background Material}
\lab{appa}
%
To design the proposed observer we rely on the following well-known fact, whose proof may be found in  \cite[Section C.4]{SONbook}.

\begin{fact} \em 
	\label{f1}
	Consider the system
	\begin{equation}
		\label{sys0}
		\dot {\mathrm z}(t) = {\mathbb A}(t) {\mathrm z}(t),  \quad {\mathrm z}(0)=z_0 \in \rea^n,
	\end{equation}
	with ${\mathrm z}(t) \in \rea^n$ and ${\mathbb A}(t) \in \rea^{n\times n}$, and the auxiliary linear $n\times n$ matrix differential equation 
	\begequ
	\lab{auxx}
	\dot {\mathbb Z}(t)= {\mathbb A}(t) {\mathbb Z}(t), \quad {\mathbb Z}(0)= I_n,
	\endequ
	with   ${\mathbb Z}(t) \in \rea^{n\times n}$. The solution of \eqref{auxx} satisfies 
	$$
	{\mathbb Z}(t)= \Phi (t)
	$$
	where $\Phi (t)$ is the state transition matrix of the system \eqref{sys0}. Consequently, the trajectories of the system \eqref{sys0} verify
	$$
	{\mathrm z}(t)= {\mathbb Z}(t){\mathrm z}_0.
	$$
	\qed
\end{fact}

%
\section*{Acknowledgment}

The authors are deeply greatful to Prof. Volker Mehrmann of TU Berlin for many helpful suggestions and corrections, which significantly improved the quality of the paper.

\section*{References}

\bibliographystyle{IEEEtran}
\bibliography{TAC_24}

\begin{thebibliography}{10}
\providecommand{\url}[1]{#1}
\csname url@rmstyle\endcsname
\providecommand{\newblock}{\relax}
\providecommand{\bibinfo}[2]{#2}
\providecommand\BIBentrySTDinterwordspacing{\spaceskip=0pt\relax}
\providecommand\BIBentryALTinterwordstretchfactor{4}
\providecommand\BIBentryALTinterwordspacing{\spaceskip=\fontdimen2\font plus
\BIBentryALTinterwordstretchfactor\fontdimen3\font minus
  \fontdimen4\font\relax}
\providecommand\BIBforeignlanguage[2]{{%
\expandafter\ifx\csname l@#1\endcsname\relax
\typeout{** WARNING: IEEEtran.bst: No hyphenation pattern has been}%
\typeout{** loaded for the language `#1'. Using the pattern for}%
\typeout{** the default language instead.}%
\else
\language=\csname l@#1\endcsname
\fi
#2}}

\bibitem{ILCREIbook}
A.~Ilchmann and T.~Reis, \emph{Surveys in Differential-Algebraic Equations
  I}.\hskip 1em plus 0.5em minus 0.4em\relax Springer Berlin, Heidelberg, 2013.

\bibitem{MEHUNG}
V.~Mehrmann and B.~Unger, ``Control of port-hamiltonian differential-algebraic
  systems and applications,'' \emph{Acta Numerica}, vol.~32, pp. 395--515,
  2023.

\bibitem{PATetalbook}
R.~J. Patton, P.~M. Frank, and R.~N. Clark, \emph{Issues of fault diagnosis for
  dynamic systems}.\hskip 1em plus 0.5em minus 0.4em\relax Springer London,
  2000.

\bibitem{BOBCAM}
K.~S. Bobinyec and S.~L. Campbell, \emph{Linear Differential Algebraic
  Equations and Observers}.\hskip 1em plus 0.5em minus 0.4em\relax Cham:
  Springer International Publishing, 2015, pp. 1--67.

\bibitem{DAIbook}
L.~Dai, \emph{Singular Control Systems}.\hskip 1em plus 0.5em minus 0.4em\relax
  Springer Berlin, Heidelberg, 1989.

\bibitem{DARscl12}
M.~Darouach, ``On the functional observers for linear descriptor systems,''
  \emph{Systems {\&} Control Letters}, vol.~61, no.~3, pp. 427--434, 2012.

\bibitem{HOUMUL}
M.~Hou and P.~Muller, ``Design of observers for linear systems with unknown
  inputs,'' \emph{IEEE Transactions on Automatic Control}, vol.~37, no.~6, pp.
  871--875, 1992.

\bibitem{MULHOU}
P.~Muller and M.~Hou, ``On the observer design for descriptor systems,''
  \emph{IEEE Transactions on Automatic Control}, vol.~38, no.~11, pp.
  1666--1671, 1993.

\bibitem{CAMDELNIK}
S.~L. Campbell, F.~Delebecque, and R.~Nikoukhah, ``Observer design for linear
  time-varying systems descriptor systems,'' \emph{Proc. Control Industrial
  Systems (CIS97), Belfort, France}, pp. 507--512, 1997.

\bibitem{ZETetal}
I.~I. Zetina-Rios, M.~Alma, G.~L. Osorio-Gordillo, M.~Darouach, and C.~M.
  Astorga-Zaragoza, ``Generalized adaptive observer design for a class of
  linear algebro-differential systems,'' in \emph{2023 IEEE 11th International
  Conference on Systems and Control (ICSC)}, 2023, pp. 171--176.

\bibitem{CAM}
S.~L. Campbell, ``Linearization of daes along trajectories,'' \emph{Zeitschrift
  f{\"u}r angewandte Mathematik und Physik ZAMP}, vol.~46, no.~1, pp. 70--84,
  1995.

\bibitem{BOB}
K.~S. Bobinyec, \emph{Observer construction for systems of differential
  algebraic equations using completions}.\hskip 1em plus 0.5em minus
  0.4em\relax North Carolina State University, 2013.

\bibitem{KUNMEHbook}
P.~Kunkel, \emph{Differential-algebraic equations: analysis and numerical
  solution}.\hskip 1em plus 0.5em minus 0.4em\relax European Mathematical
  Society, 2006, vol.~2.

\bibitem{ORTetalaut}
R.~Ortega, A.~Bobtsov, N.~Nikolaev, J.~Schiffer, and D.~Dochain, ``Generalized
  parameter estimation-based observers: Application to power systems and
  chemical–biological reactors,'' \emph{Automatica}, vol. 129, p. 109635,
  2021.

\bibitem{BEZetal}
V.~Bezzubov, A.~Bobtsov, D.~Efimov, R.~Ortega, and N.~Nikolaev, ``Adaptive
  state observation of linear time-varying systems with delayed measurements
  and unknown parameters,'' \emph{International Journal of Robust and Nonlinear
  Control}, vol.~33, no.~2, pp. 1203--1213, 2023.

\bibitem{BOBetalijc}
B.~Y. Alexey~Bobtsov, Romeo~Ortega and N.~Nikolaev, ``Adaptive state estimation
  of state-affine systems with unknown time-varying parameters,''
  \emph{International Journal of Control}, vol.~95, no.~9, pp. 2460--2472,
  2022.

\bibitem{BOBetalaut}
A.~Bobtsov, N.~Nikolaev, R.~Ortega, and D.~Efimov, ``State observation of {LTV}
  systems with delayed measurements: A parameter estimation-based approach with
  fixed convergence time,'' \emph{Automatica}, vol. 131, p. 109674, 2021.

\bibitem{PYRetal}
A.~Pyrkin, A.~Bobtsov, R.~Ortega, and A.~Isidori, ``An adaptive observer for
  uncertain linear time-varying systems with unknown additive perturbations,''
  \emph{Automatica}, vol. 147, p. 110677, 2023.

\bibitem{ROMORT}
J.~G. Romero and R.~Ortega, ``Adaptive state observation of linear time-varying
  systems with switching unknown parameters: Application to gain scheduling and
  event-triggered control,'' \emph{International Journal of Adaptive Control
  and Signal Processing}, vol.~37, no.~11, pp. 2915--2933, 2023.

\bibitem{WANORTBOB}
L.~Wang, R.~Ortega, and A.~Bobtsov, ``Observability is sufficient for the
  design of globally exponentially stable state observers for state-affine
  nonlinear systems,'' \emph{Automatica}, vol. 149, p. 110838, 2023.

\bibitem{BERbook}
P.~Bernard, \emph{Observer design for nonlinear systems}.\hskip 1em plus 0.5em
  minus 0.4em\relax Springer, 2019, vol. 479.

\bibitem{RUGbook}
W.~J. Rugh, \emph{Linear system theory (2nd ed.)}.\hskip 1em plus 0.5em minus
  0.4em\relax Prentice-Hall, Inc., 1996.

\bibitem{ILC}
A.~ILCHMANN, \emph{{Contributions to Time-Varying Linear Control
  Systems}}.\hskip 1em plus 0.5em minus 0.4em\relax Verlag an der Lottbek,
  1989.

\bibitem{CAMPET}
S.~L. Campbell and L.~R. Petzold, ``Canonical forms and solvable singular
  systems of differential equations,'' \emph{SIAM Journal on Algebraic Discrete
  Methods}, vol.~4, no.~4, pp. 517--521, 1983.

\bibitem{BunBMN99}
A.~Bunse-Gerstner, R.~Byers, V.~Mehrmann, and N.~K. Nichols, ``Feedback design
  for regularizing descriptor systems,'' \emph{Linear algebra and its
  applications}, vol. 299, no. 1-3, pp. 119--151, 1999.

\bibitem{CamM09}
S.~L. Campbell and C.~D. Meyer, \emph{Generalized inverses of linear
  transformations}.\hskip 1em plus 0.5em minus 0.4em\relax SIAM, 2009.

\bibitem{SASBODbook}
S.~Sastry and M.~Bodson, \emph{Adaptive control: stability, convergence and
  robustness}.\hskip 1em plus 0.5em minus 0.4em\relax Prentice-Hall, New
  Jersey, 1989.

\bibitem{TAObook}
G.~Tao, \emph{Adaptive Control Design and Analysis}, ser. Adaptive and
  Cognitive Dynamic Systems: Signal Processing, Learning, Communications and
  Control.\hskip 1em plus 0.5em minus 0.4em\relax Wiley, 2003.

\bibitem{ORTROMARA}
R.~Ortega, J.~G. Romero, and S.~Aranovskiy, ``A new least squares parameter
  estimator for nonlinear regression equations with relaxed excitation
  conditions and forgetting factor,'' \emph{Systems {\&} Control Letters}, vol.
  169, p. 105377, 2022.

\bibitem{KRERIE}
G.~Kreisselmeier and G.~Rietze-Augst, ``Richness and excitation on an
  interval-with application to continuous-time adaptive control,'' \emph{IEEE
  Transactions on Automatic Control}, vol.~35, no.~2, pp. 165--171, 1990.

\bibitem{KunMR01}
P.~Kunkel, V.~Mehrmann, and W.~Rath, ``Analysis and numerical solution of
  control problems in descriptor form,'' \emph{Mathematics of Control, Signals
  and Systems}, vol.~14, pp. 29--61, 2001.

\bibitem{ORTetaltac}
R.~Ortega, S.~Aranovskiy, A.~A. Pyrkin, A.~Astolfi, and A.~A. Bobtsov, ``New
  results on parameter estimation via dynamic regressor extension and mixing:
  Continuous and discrete-time cases,'' \emph{IEEE Transactions on Automatic
  Control}, vol.~66, no.~5, pp. 2265--2272, 2021.

\bibitem{ORTNIKGER}
R.~Ortega, V.~Nikiforov, and D.~Gerasimov, ``On modified parameter estimators
  for identification and adaptive control. a unified framework and some new
  schemes,'' \emph{Annual Reviews in Control}, vol.~50, pp. 278--293, 2020.

\bibitem{TRE}
S.~Trenn, \emph{Solution Concepts for Linear DAEs: A Survey}.\hskip 1em plus
  0.5em minus 0.4em\relax Berlin, Heidelberg: Springer Berlin Heidelberg, 2013,
  pp. 137--172.

\bibitem{TRAetal}
M.~Tranninger, H.~Niederwieser, R.~Seeber, and M.~Horn, ``Unknown input
  observer design for linear time-invariant systems—a unifying framework,''
  \emph{International Journal of Robust and Nonlinear Control}, vol.~33,
  no.~15, pp. 8911--8934, 2023.

\bibitem{DAVFRILEV}
J.~D{\'a}vila, L.~Fridman, and A.~Levant, ``Robust state estimation for linear
  time-varying systems using high-order sliding-modes observers,'' in
  \emph{Sliding-Mode Control and Variable-Structure Systems: The State of the
  Art}.\hskip 1em plus 0.5em minus 0.4em\relax Springer, 2023, pp. 133--162.

\bibitem{SONbook}
E.~D. Sontag, \emph{Mathematical control theory: deterministic finite
  dimensional systems}.\hskip 1em plus 0.5em minus 0.4em\relax Springer Science
  \& Business Media, 2013, vol.~6.

\end{thebibliography}

\begin{IEEEbiography}[{\includegraphics[width=1in,height=1.25in,clip,keepaspectratio]{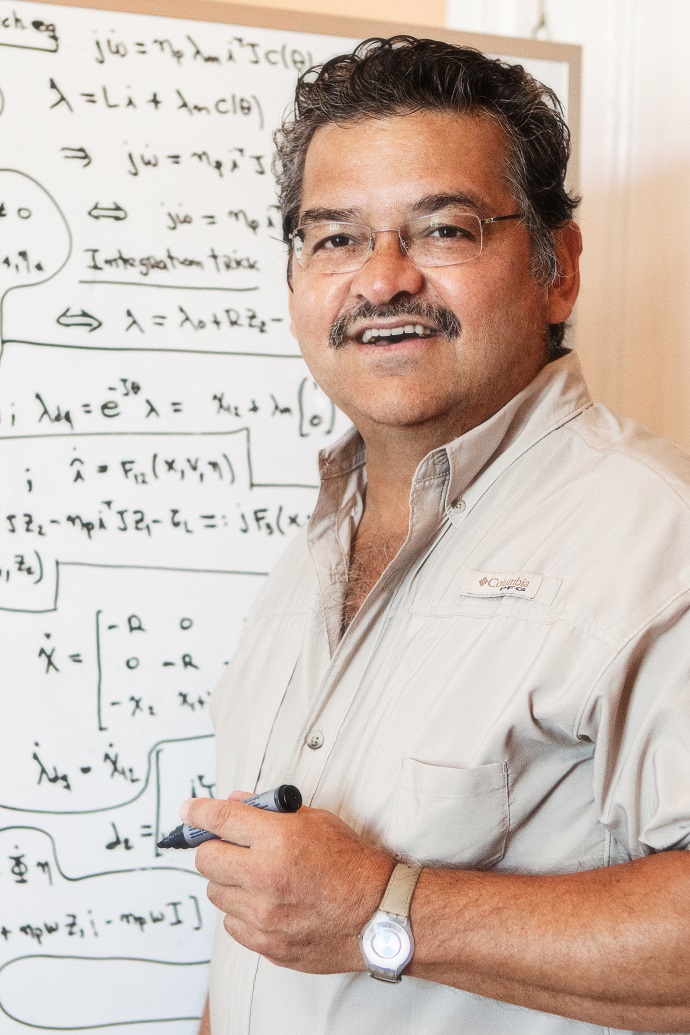}}]{Romeo Ortega} (Fellow, IEEE) was born in Mexico. He obtained his B.S. in Electrical and Mechanical Engineering from the National University of Mexico, Master of Engineering from Polytechnical Institute of Leningrad, USSR, and the Docteur d‘Etat from the Polytechnical Institute of Grenoble, France in 1974, 1978 and 1984, respectively.
	
	He then joined the National University of Mexico, where he worked until 1989. He was a Visiting Professor at the University of Illinois in 1987–1988 and at the McGill University in 1991–1992, and a Fellow of the Japan Society for Promotion of Science in 1990–1991. He has been a member of the French National Researcher Council (CNRS) since June 1992. Currently, he is in the Laboratoire de Signaux et Systemes (SUPELEC) in Paris. His research interests are in the fields of nonlinear and adaptive control, with special emphasis on applications.
	
	Dr. Ortega has published three books and more than 350 scientific papers in international journals, with an h-index of 97. He has supervised more than 35 Ph.D. theses. He is a Fellow Member of the IEEE since 1999 and an IFAC fellow since 2016. He has served as Chairman in several IFAC and IEEE committees and participated in various editorial boards of international journals. Currently he is Editor in Chief of Int. J. Adaptive Control and Senior Editor of Asian J. of Control.
\end{IEEEbiography}

\begin{IEEEbiography}[{\includegraphics[width=1in,height=1.25in,clip,keepaspectratio]{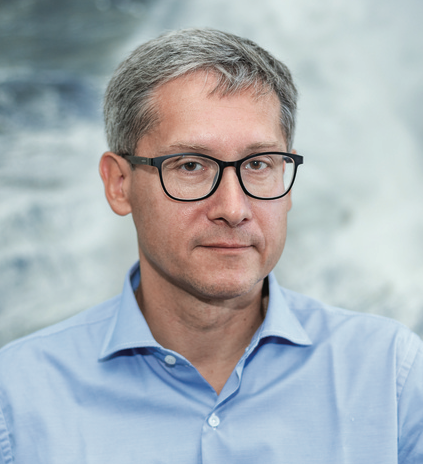}}]{Alexey Bobtsov} received the M.S. degree in Electrical Engineering from ITMO University, St. Petersburg, Russia in 1996, received his Ph.D. in 1999 and the degree of Doctor of Science (habilitation thesis) in 2007 from the same University. From December 2000 to May 2007 Dr. Bobtsov served as Associate Professor of Department of Control Systems and Informatics. In May 2007 Dr. Bobtsov was appointed as Professor of Department of Control Systems and Informatics. In September 2008 he was elected as the Dean of Computer Technologies and Controlling Systems Faculty. He is currently a Director of School of Computer Technologies and Control at ITMO University. He is a Senior Member of IEEE since 2010. He is a Member of International Public Association Academy of Navigation and Motion Control. He is a co-author of more than 250 journal and conference papers, 20 books and textbooks. His research interests are in fields of nonlinear and adaptive control.
\end{IEEEbiography}

\begin{IEEEbiography}[{\includegraphics[width=1in,height=1.25in,clip,keepaspectratio]{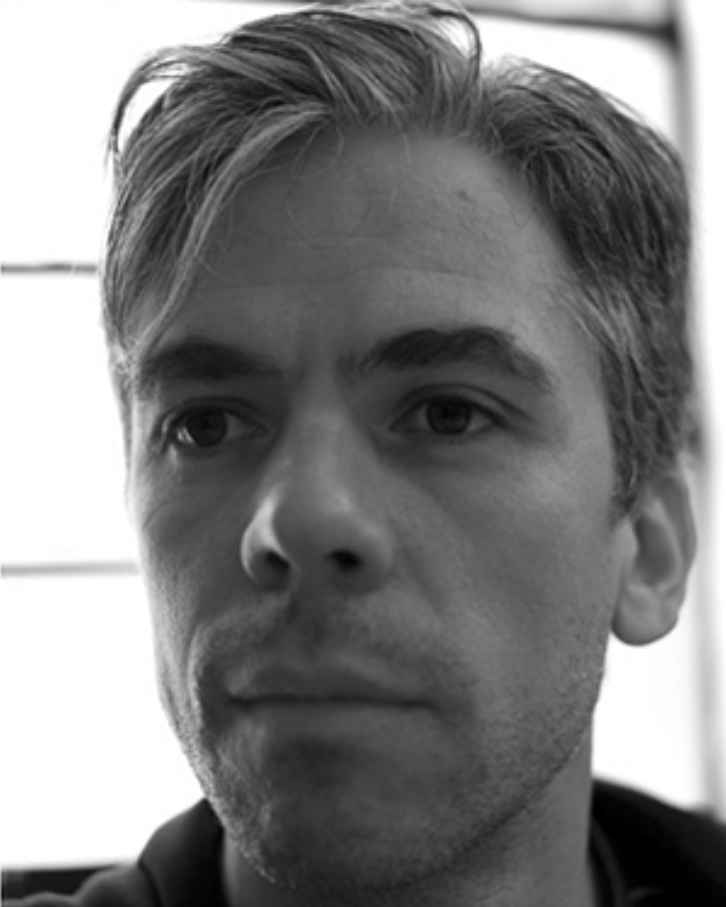}}]{Fernando Castaños} received the B.Eng. degree in electric and electronic engineering and the M.Eng. degree in control engineering from the Universidad Nacional Autónoma de México, Mexico City, Mexico, in 2002 and 2004, respectively, and the Ph.D. degree in control theory from Université Paris-Sud XI, Paris, France, in 2009.
	
	He was a Postdoctoral Fellow with the Center for Intelligent Machines, McGill University, Montréal, QC, Canada, for two years. Since 2011, he has been with the Departamento de Control Automático, CINVESTAV-IPN, Mexico City. His research interests include variable structure systems, passivity-based control, nonlinear control, port-Hamiltonian systems, and robust control.
	Dr. Castaños is an Editor for International Journal of Robust and Nonlinear Control.
	
\end{IEEEbiography}

\begin{IEEEbiography}[{\includegraphics[width=1in,height=1.25in,clip,keepaspectratio]{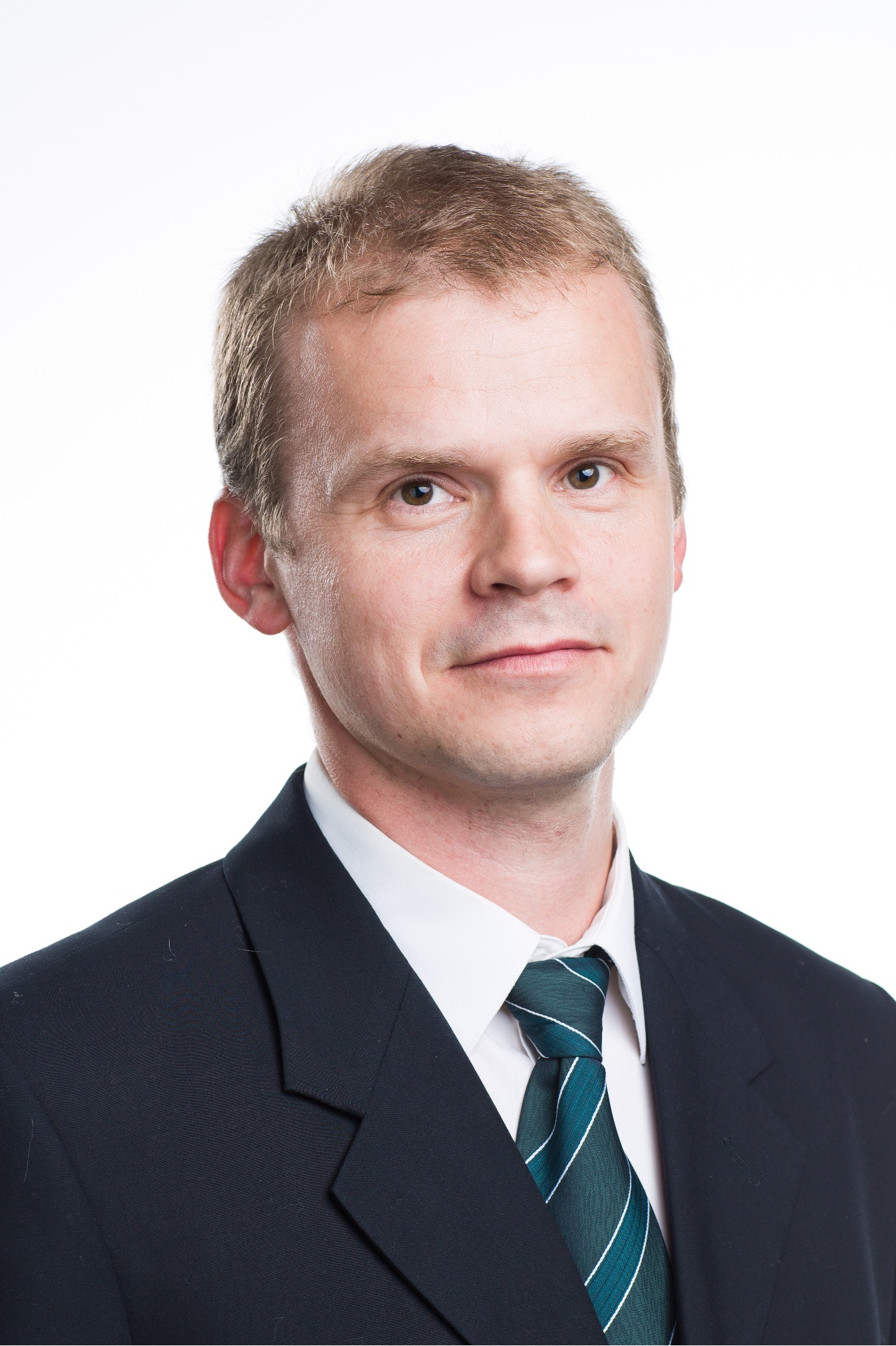}}]{Nikolay Nikolaev} received the M.S. degree in electrical engineering from ITMO University, St. Petersburg, Russia in 2003, received his PhD in 2006 from the same University. From 2002 until 2013 he worked as Engineer of Department of Control Systems Design for Power Plants at JSC Kirovsky Zavod (Kirov Plant). From 2013 he is an Associate Professor in Department of Control Systems and Robotics from ITMO University. He is a Member of IEEE since 2006. His research interests are in fields of nonlinear and adaptive control.
\end{IEEEbiography}

\end{document}